\RequirePackage{fix-cm}
\documentclass[twocolumn,epjc3]{svjour3}  
\smartqed  
\RequirePackage{graphicx}
\journalname{Eur. Phys. J. C}
\usepackage{amsmath}
\usepackage{verbatim}
\usepackage{url}
\usepackage[utf8]{inputenc}
\usepackage[english]{babel}
\usepackage[T1]{fontenc}

\usepackage{amssymb}
\usepackage{amsfonts}
\usepackage{textcomp}
\usepackage[export]{adjustbox}
\graphicspath{{./Figure}}
\usepackage{url}
\usepackage{hyperref}

\begin{document}

\title{\boldmath High-rate tests on Resistive Plate Chambers operated with eco-friendly gas mixtures}

\author{
\textbf{The RPC ECOGas@GIF++ Collaboration:}\\
M. Abbrescia \thanksref{BariUniv, INFNBari, e1}\and
G. Aielli\thanksref{TorVergataUniv, INFNTorVergata}\and
R. Aly\thanksref{INFNBari, HelwanUniv}\and
M. C. Arena\thanksref{PaviaUniv}\and
M. Barroso\thanksref{RioUniv}\and
L. Benussi\thanksref{INFNFrascati}\and
S. Bianco\thanksref{INFNFrascati}\and
F. Bordon\thanksref{CERN} \and
D. Boscherini\thanksref{INFNBologna}\and
A. Bruni\thanksref{INFNBologna}\and
S. Buontempo\thanksref{INFNNapoli}\and
M. Busato\thanksref{CERN}\and
P. Camarri\thanksref{TorVergataUniv, INFNTorVergata}\and
R. Cardarelli\thanksref{INFNTorVergata}\and
L. Congedo\thanksref{BariUniv, INFNBari}\and
D. De Jesus Damiao\thanksref{RioUniv}\and
M. De Serio\thanksref{BariUniv, INFNBari}\and
A. Di Ciaccio\thanksref{TorVergataUniv, INFNTorVergata}\and
L. Di Stante\thanksref{TorVergataUniv, INFNTorVergata}\and
P. Dupieux\thanksref{Clermont}\and
J. Eysermans\thanksref{MIT}\and
A. Ferretti\thanksref{TorinoUniv, INFNTorino}\and
G. Galati\thanksref{BariUniv, INFNBari}\and
M. Gagliardi\thanksref{TorinoUniv, INFNTorino}\and
R. Guida\thanksref{CERN}\and
G. Iaselli\thanksref{INFNBari, PoliBariUniv}\and
B. Joly\thanksref{Clermont}\and
K. S. Lee\thanksref{SeoulUniv}\and
B. Liberti\thanksref{INFNTorVergata}\and
D. Lucero Ramirez\thanksref{MexicoCityUniv}\and 
B. Mandelli\thanksref{CERN}\and
S.P. Manen\thanksref{Clermont}\and
L. Massa\thanksref{INFNBologna}\and
A. Pastore\thanksref{INFNBari}\and
E. Pastori\thanksref{INFNTorVergata}\and
D. Piccolo\thanksref{INFNFrascati}\and
L. Pizzimento\thanksref{INFNTorVergata}\and
A. Polini\thanksref{INFNBologna}\and
G. Proto\thanksref{TorVergataUniv, INFNTorVergata}\and
G. Pugliese\thanksref{INFNBari, PoliBariUniv}\and
L. Quaglia\thanksref{INFNTorino}\and
D. Ramos\thanksref{INFNBari, PoliBariUniv}\and
G. Rigoletti\thanksref{CERN}\and
A. Rocchi\thanksref{INFNTorVergata}\and
M. Romano\thanksref{INFNBologna}\and
A. Samalan\thanksref{GentUniv}\and
P. Salvini\thanksref{INFNPavia}\and
R. Santonico\thanksref{TorVergataUniv, INFNTorVergata}\and
G. Saviano\thanksref{SapienzaUniv}\and
S. Simone\thanksref{BariUniv, INFNBari}\and
L. Terlizzi\thanksref{TorinoUniv, INFNTorino}\and
M. Tytgat\thanksref{GentUniv, VrijeUniv}\and
E. Vercellin\thanksref{TorinoUniv, INFNTorino}\and
M. Verzeroli\thanksref{CERN, LyonUniv}\and
N. Zaganidis\thanksref{MexicoCityUniv}}

\institute{
Universit\`a of Bari, Dipartimento Interateneo di Fisica, via Amendola 173, 70125 Bari, Italy\label{BariUniv}\and
INFN Sezione di Bari, Via E. Orabona 4, 70125 Bari, Italy\label{INFNBari}\and
Universit\`a degli studi di Roma Tor Vergata, Dipartimento di Fisica, Via della Ricerca Scientifica 1, 00133 Roma, Italy\label{TorVergataUniv}\and
INFN Sezione di Roma Tor Vergata, Via della Ricerca Scientifica 1, 00133 Roma, Italy\label{INFNTorVergata}\and
Helwan University, Helwan Sharkeya, Helwan, Cairo Governorate 4037120, Egypt\label{HelwanUniv}\and
Universit\`a degli studi di Pavia, Corso Strada Nuova 65, 27100 Pavia, Italy\label{PaviaUniv}\and
Universidade do Estado do Rio de Janeiro, R. Sao Francisco Xavier, 524 - Maracan\'a, Rio de Janeiro - RJ, 20550-013, Brasil\label{RioUniv}\and
INFN - Laboratori Nazionali di Frascati, Via Enrico Fermi 54, 00044 Frascati (Roma), Italy\label{INFNFrascati}\and
CERN, Espl. des Particules 1, 1211 Meyrin, Svizzera\label{CERN}\and
INFN Sezione di Bologna, Viale C. Berti Pichat 4/2, 40127 Bologna, Italy\label{INFNBologna}\and
INFN Sezione di Napoli, Complesso Universitario di Monte S. Angelo ed. 6, Via Cintia, 80126 Napoli, Italy\label{INFNNapoli}\and
Clermont Universit\'e, Universit\'e Blaise Pascal, CNRS/IN2P3, Laboratoire de Physique Corpusculaire, BP 10448, F-63000 Clermont-Ferrand, France\label{Clermont}\and
Massachusetts Institute of Technology, 77 Massachusetts Ave, Cambridge, MA 02139, USA\label{MIT}\and
Universit\`a degli studi di Torino, Dipartimento di Fisica, Via P. Giuria 1, 10126 Torino, Italy\label{TorinoUniv}\and
INFN Sezione di Torino, Via P. Giuria 1, 10126 Torino, Italy\label{INFNTorino}\and
Politecnico di Bari, Dipartimento Interateneo di Fisica, via Amendola 173, 70125 Bari, Italy\label{PoliBariUniv}\and
Korea University, Departiment of Physics, Seoul\label{SeoulUniv}\and
Universidad Iberoamericana, Dept. de Fisica y Matematicas, Mexico City 01210, Mexico\label{MexicoCityUniv}\and
Ghent University, Dept. of Physics and Astronomy, Proeftuinstraat 86, B-9000 Ghent, Belgium\label{GentUniv}\and
INFN Sezione di Pavia, Via A. Bassi 6, 27100 Pavia, Italy\label{INFNPavia}\and
Universit\`a di Roma "Sapienza", Dipartimento di Ingegneria Chimica Materiali Ambiente, Piazzale Aldo Moro 5, 00185 Roma, Italy\label{SapienzaUniv}\and
Vrije Universiteit Brussel (VUB-ELEM), Dept. of Physics, Pleinlaan 2, 1050 Brussels, Belgium\label{VrijeUniv}\and
Universit\'e Claude Bernard Lyon I, 43 Bd du 11 Novembre 1918, 69100 Villeurbanne, France\label{LyonUniv}
}




\thankstext[$^\star$]{e1}{Corresponding author: Marcello Abbrescia, e-mail: marcello.abbrescia@ba.infn.it}



\date{Received: date / Accepted: date}

\maketitle

\begin{abstract}
Results obtained by the RPC ECOgas@GIF++ Collaboration, using Resistive Plate Chambers operated with new, eco-friendly gas mixtures, based on Tetrafluoropropene and carbon dioxide, are shown and discussed in this paper. Tests aimed to assess the performance of this kind of detectors in high-irradiation conditions, analogous to the ones foreseen for the coming years at the Large Hadron Collider experiments, were performed, and demonstrate a performance basically similar to the one obtained with the gas mixtures currently in use, based on Tetrafluoroethane, which is being progressively phased out for its possible contribution to the greenhouse effect. Long term aging tests are also being carried out, with the goal to demonstrate the possibility of using these eco-friendly gas mixtures during the whole High Luminosity phase of the Large Hadron Collider.
\end{abstract}

\keywords{Resistive Plate Chambers \and Gaseous detectors \and Eco-friendly gas mixtures \and GIF++}

%

\PACS{29.40.Cs}

\section{Introduction}
\label{sec:intro}

Resistive Plate Chambers (RPCs) are a widespread detector, used both for astro-particle and accelerator physics experiments, as well as environmental monitoring and medical applications. In particular, at the Large Hadron Collider (LHC) at CERN, all four big experiments, namely ALICE, ATLAS, CMS and LHCb, use, or plan to use, RPCs, in slightly different configurations and for multiple purposes  \cite{ALICE1} \cite {ATLAS1} \cite{CMS1} \cite {LHCb1}. The RPCs of the LHC experiments have been in operation for more than ten years now, contributing to some of the most important discoveries in particle physics in the last dozen years.

Presently, the RPCs used at the LHC experiments are operated with gas mixtures with Tetrafluoroethane (TFE) as main component. TFE, whose empirical formula is C$_2$H$_2$F$_4$ and is also commercially known as R-134a, is usually mixed with iso-butane (i-C$_4$H$_{10}$) and Sulfur hexafluoride (SF$_6$) in various percentages, to exploit the well-known discharge quenching properties of these gases. However, TFE is characterized by a Global Warming Potential (GWP\footnote{The GWP of a certain gas depends on the time frame considered. Here and in the following we will refer to the GWP over 100 years. For all tables and calculations, moreover, we will use the GWP reported in \cite{Europe}.}) around 1430, namely it contributes to the greenhouse effect 1430 times more than an equivalent mass of CO$_2$, and regulations by the European Community, derived from the adoption of the Kyoto protocol, prohibit it for many applications \cite{Europe}. Even if scientific applications do not fall within these restrictions, the collaborations of the LHC experiments decided to investigate for possible replacements of TFE with other, more eco-friendly, gases.

Studies with the goal of finding a new eco-friendly gaseous component suitable to replace TFE in RPCs  started several years ago. Given the fact that the present RPC systems of the experiments at LHC feature several cubic meters of gas volume, this search could not consider the whole set of eco-friendly gases, but only the ones that are available on the market at a reasonable cost. Summarizing the work of many years, the main idea was to replace TFE with tetrafluoropropene, whose empirical formula is C$_3$H$_2$F$_4$, which is an  hydrofluoroolefin (HFO) characterized by a quite low GWP. In particular, its allotropic form, commercially known as HFO-1234ze, proved to be the most suitable for applications in detectors for particle physics, and is characterized by a GWP around 7 \cite{Europe}. HFO-1234ze is widely used in refrigeration industry, making it easily to procure and relatively not expensive.

HFO-1234ze has a molecule quite similar to TFE, nevertheless its first effective Townsend coefficient, at a given electric field strength, is lower with respect to the one of TFE (for a recent measurement of gas parameters of HFO-1234ze and a comparison with TFE, see \cite{Fan}). In the present-day RPCs, replacing TFE with HFO-1234ze would result in too large operating voltages to be compatible with the high voltage systems and readout electronics used at the LHC experiments. Therefore, in order to keep the operating voltage within an acceptable range, it was proposed to replace TFE with a binary mixture, made either of HFO-1234ze and CO$_2$, or HFO-1234ze and He. However, He cannot be used at collider experiments, for problems it might cause to photomultipliers in some other experimental subsystems (see, for instance, \cite{Photo1} \cite{Photo2}), therefore the studies concentrated mainly on mixtures made out of HFO-1234ze and CO$_2$, in various percentages.

Note that it has been pointed out that HFO-1234ze, in the high atmosphere, can dissociate giving eventually origin to trifluoroacetic acid (TFA), which is potentially harmful for the environment and the human health, if removed from the atmosphere by rainfalls \cite{TFA1} \cite{TFA2}. A long debate has been going on the subject, whose outcome seems to demonstrate that the actual impact should be irrelevant (see \cite{TFA3} and references therein). Nevertheless, this is one of the potential issues to be considered, and will probably require deeper insights in the future.

Some studies about the performance of RPCs filled with HFO-1234ze/CO$_2$ mixtures have already been reported, showing encouraging results, in terms of performance obtained \cite{ALICE2} \cite{ATLAS2} \cite{CMS2} \cite{EP-DT} \cite{LHCb2}. Nevertheless, at the moment, a long term aging test of RPCs filled with eco-friendly gas mixtures based on HFO-1234ze and CO$_2$, operated under a high particle radiation background, is still missing, and this is a crucial point if such gas mixtures are to be considered for use during the High-Luminosity phase of the LHC (HL-LHC).

Joined by the common interest for this topic, a collaboration across multiple groups working on RPCs at the LHC experiments was established, with the specific goal to carry out these long term, high irradiation conditions, tests. 
Groups from the ALICE, ATLAS, CMS, LHCb/SHiP experiments, together with the EP-DT gas group from CERN, are components of this collaboration, usually called the "RPC ECOGas@GIF++ Collaboration"\footnote{Later on, activities related to the search of eco-friendly gas mixtures for RPCs were also carried out in the framework of the Working Group 7.2 of the AidaInnova project \cite{AidaInnova}.}.

Testing the very same HFO-1234ze/CO$_2$ gas mixtures, with the different detector layouts and front-end electronics brought by the various groups, provides deep insights both on the gas and the detector behaviours, and, in principle, may allow to disentangle the various effects that could be related either to the specific designs and/or to the production techniques, and/or the electronics used.

\section{Experimental set-up}
\label{sec:description}

The results presented in this paper were obtained during several beam tests, which took place from July until October 2021, using the H4-SPS secondary beam line available at the Gamma Irradiation Facility at CERN  \cite{GIF++} (usually simply referred to as GIF++). Here detectors can be kept under irradiation with photons from a 12.5 TBq $^{137}$Cs source, simulating the irradiation conditions foreseen during the HL-LHC, while measuring their performance using the beam. Usually a 100 GeV muon beam was used for this purpose \cite{Beam}. A system of five absorption filters (ABS), made in Lead, mounted on movable supports, can be used to adjust the intensity of the radiation field produced by the $^{137}$Cs source, namely the rate of photons impinging onto the detectors, down to a factor 50000 from its maximum intensity. An additional Aluminium filter, positioned close to the source and suitably shaped, is also used to make the $\gamma$ flux as uniform as possible in vertical planes perpendicular to the muon beam and the bunker walls; this is a useful feature to uniformly irradiate large planar detectors.

\begin{figure*}[tbp]
\centering 
\includegraphics[width=0.7\linewidth]{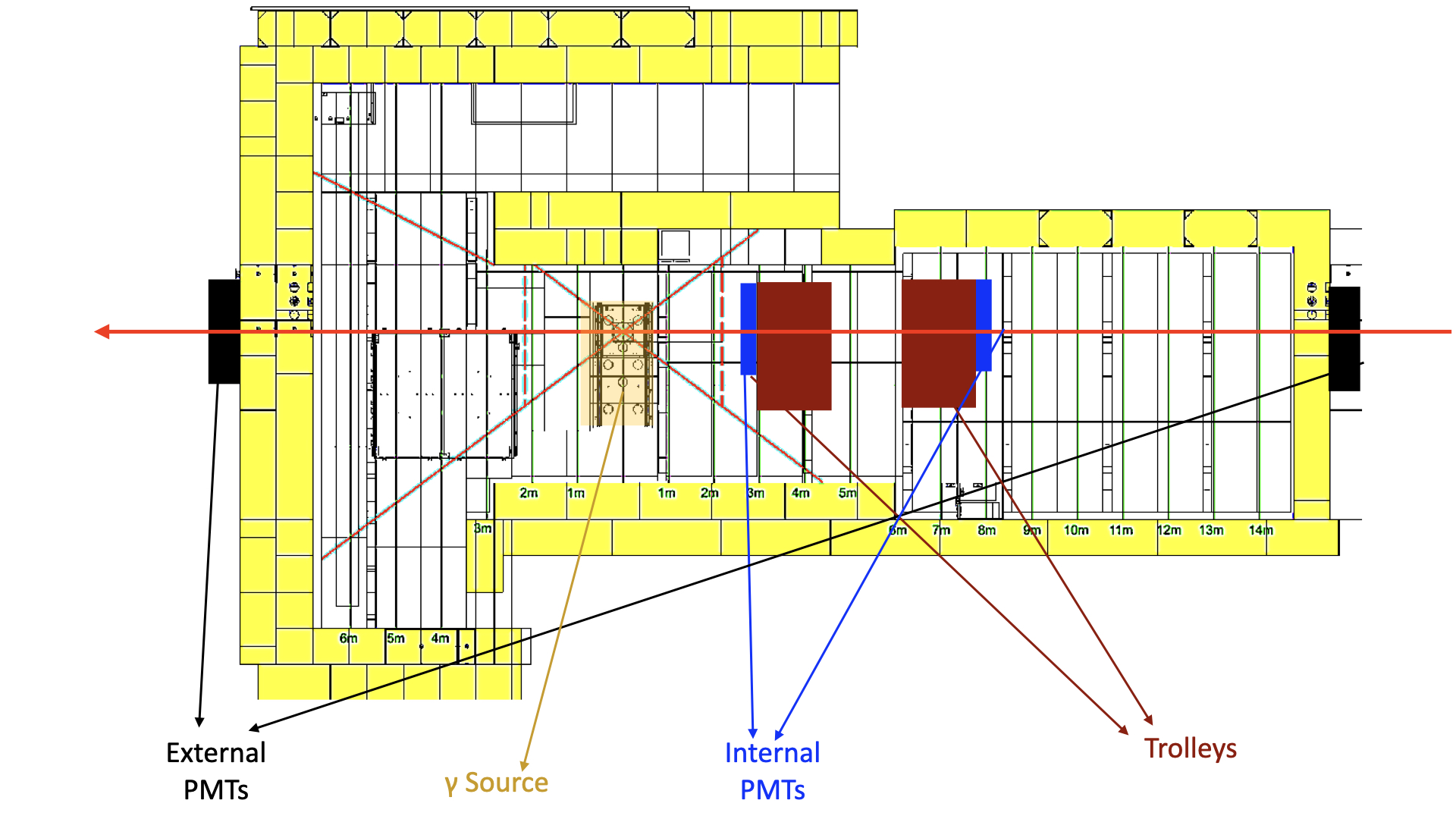}
\caption{Layout of the GIF++ facility, indicating: the position of the $^{137}$Cs source (light brown), the positions of the two trolleys hosting the chambers under test (dark brown), the positions of the scintillators internal to the GIF++ (blue), the positions of the scintillators external to the GIF++ (black). The red arrow indicates the muon beam from the H4 line used for the tests, and the yellows blocks the concrete walls. The grey lines approximately delimit the irradiation cone from the $^{137}$Cs source.}
\label{fig:GIF++}
\end{figure*}

Inside the GIF++, the detectors to be tested were mounted onto two mechanical support structures, positioned at around 3 and 6 m from the $^{137}$Cs source. In particular, chambers from ATLAS, CERN EP-DT and CMS groups were mounted on a trolley positioned 3 m away from the source, while chambers from ALICE and LHCb/SHiP on a trolley 6 m away. A simplified sketch of the GIF++ layout, re-elaborated from \cite{GIF++2},  with, highlighted, the position of the two trolleys, as well as other features, is shown in Figure~\ref{fig:GIF++}. The characteristics of the chambers under test, in terms of distance from the source, dimensions, gap size, electrode thickness, number of strips, strip pitch, are listed in Table \ref{tab:chamberdata}. Note that the chambers from ALICE, ATLAS, CERN EP-DT and LHCb/SHiP groups were rectangular in shape, and single-gap, while the chamber from CMS was double-gap and trapezoidal in shape, equal to the ones actually used in the endcaps of the experiment \cite{CMS1}. Also the chamber from ALICE was of the same kind of the ones currently in use \cite{ALICE1} in the experiment, while the others were prototypes built for these and other performance studies. 

For signal readout purposes, the chambers under test were equipped with sets of copper strips. While the chambers from the CMS and CERN EP-DT groups featured one set of strips, horizontally directed, chambers from ALICE and LHCb/SHiP had two sets of strips, perpendicular to each other, positioned on the opposite sides of the gas gaps. The chamber from ATLAS was equipped with one single central strip, 3 cm wide, while the trapezoidal shape of the CMS chamber resulted in a variable strip pitch.

\begin{table*}
\caption{Characteristics of the chambers under test. All chambers were rectangular, except the one from CMS, which was trapezoidal in shape, so dimensions of the smaller and larger bases are shown, as well as the minimum and maximum strip pitch. The ALICE and LHCb/SHiP detectors were equipped with two sets of strips, perpendicular to each other, while the ATLAS chamber with one single central strip, 3 cm wide. In the ALICE case the readout strips covered basically the entire detector surface, while for EP-DT and LHCb/SHiP they were positioned just onto their central part.}
\vspace{0.5cm}
\centering
\label{tab:chamberdata}
  \begin{tabular}{ | p{1.8cm} |p{1.5cm} |p{2.0cm} |p{1.5cm} |p{1.9cm}  | p{1.7cm}| p{1.7cm}| }
\hline
  Group & Distance from the $^{137}$Cs (m) & Dimensions (cm $\times$ cm) & Gap thickness (mm) & Electrode thickness (mm) &  \# of strips & Strip pitch (mm) \\ \hline  
{\em ALICE} & 6.0 & 50 $\times$ 50 & 2 & 2 & 16 + 16 & 30\\ \hline
{\em ATLAS} & 3.2 & 55 $\times$ 10 & 2 & 1.8 & 1 & N/A \\ \hline
{\em CMS} & 3.5 & (41.5 $\div$ 23.9) $\times$ 100.5 & 2 + 2 & 2 & 128 & 5.5 $\div$ 11\\ \hline
{\em EP-DT} & 3.0 & 70 $\times$ 100 & 2 & 2 & 7 & 21\\ \hline
{\em LHCb/SHiP} & 6.2 & 70 $\times$ 100 & 1.6 & 1.6 & 32+32 & 10.6\\ 
\hline
\end{tabular}
\end{table*}

In the ALICE, CMS and LHCb/SHiP chambers, signals induced on the readout strips were amplified, discriminated and formed by means of suitable front-end electronics which was connected to Time to Digital Converters (TDCs), downstream in the Data Acquisition (DAQ) chain. ATLAS and the CERN EP-DT groups used suitable digitizers, which directly acquired the wave-forms from the readout strips and stored them for subsequent analysis. 
Each experiment analyzed the data, either from TDCs or from the digitizers, with its own custom analysis algorithms. Some characteristics of the readout electronics used, namely the Front-End (FE) electronics chip type, FE equivalent threshold, DAQ (TDC or Digitizer), are reported in Table \ref{tab:electronics}. Note that the front-end electronics used for these tests is also currently used in two of the RPC systems at LHC, namely the FFERIC chip in the ALICE RPC system, and the electronics mounted on the CMS (for details on the front-end electronics employed, see \cite{FEERIC} \cite{CMSRPC}).

Throughout the data acquisition, a custom made Detector Control Software (DCS) package collected information about the environmental parameters in the GIF++ bunker, as well as the state of each detector, and stored them for later analysis \cite{GIF++}. During the performance tests using the muon beam, the trigger was obtained using the coincidence of the signals from four scintillators, two of them located outside the irradiation zone and two inside it, defining by their intersection a 10 $\times$ 10 cm$^2$ area. 

\begin{table*}
    \caption{Some characteristics of the electronics used for readout of the various chambers under test. The ALICE, CMS and LHCb/SHiP chambers were equipped with frontend electronics, featuring signals amplification, therefore the threshold after amplification is indicated here.  The ALICE and LHCb/SHiP chambers featured two sets of readout strips, positioned in the horizontal and  vertical directions, therefore both thresholds are indicated. The ATLAS and CERN EP-DT were readout by means of suitable digitizers.}
\vspace{0.5cm}
\centering
\label{tab:electronics}
  \begin{tabular}{|p{2.0cm} |p{2.5cm} |p{6.0cm} |p{3.0cm} | }
\hline
  Group & FE chip & FE equivalent threshold & DAQ type \\ \hline  
{\em ALICE} & FEERIC ASICs & -200 mV for horizontal strips, 106 mV vertical strips (after amplification) & TDC CAEN mod.V1190A\\ \hline
{\em ATLAS} & N/A & 130 fC & Digitizer CAEN mod.DT5730\\ \hline
{\em CMS} & CMS/RPC standard electronics & 220 mV = 150 fC (after amplification) & TDC CAEN mod.V1190A\\ \hline
{\em EP-DT} & N/A & 2 mV & Digitizer CAEN mod.V1730\\ \hline
{\em LHCb/SHiP} & FEERIC ASICs & - 70 mV for horizontal strips, 80 mV for vertical strips (after amplification) & TDC CAEN mod.V1190A\\ 
\hline
\end{tabular}
\end{table*}

The tests, whose results are reported in this paper, were performed when flushing the RPCs with the so called "standard" gas mixture (STD mixture in the following), used in the RPCs of the CMS experiment, whose main component, as already pointed out, is TFE. Subsequently, the RPCs were filled with gas mixtures where TFE was replaced by HFO-1234ze and CO$_2$, in different percentages. We tested, in particular, two gas mixtures, hereafter conventionally called ECO2 and ECO3 \footnote{The ECO1 gas mixture was studied in an earlier phase of the tests on eco-friendly gas mixtures performed by the RPC ECOGas@GIF++ collaboration, and featured a larger content of HFO-1234ze. It was soon discarded because first tests indicated a too large operating voltage and an increase, with time, of the current absorbed by the detectors, maybe associated to an augmented production of impurities in the gas volume, with respect to the STD gas mixture \cite{Rigol}.}. The percentage compositions, in volume, of the gas mixtures used, and their GWP, are listed in Table \ref{tab:gasmixutre}. 
The ECO2 and ECO3 gas mixtures were designed to reduce the greenhouse gases emissions with respect to the STD gas mixture, and in fact they feature, respectively, a GWP of around 476 and 527, namely, roughly a factor three less than the STD mixture.  

Note that the GWP of the gas mixtures, listed in Table \ref{tab:gasmixutre}, are computed, as prescribed in \cite{Europe}, as the weighted average of the GWPs of the respective gaseous components, where the weights used for the average are their fractions in mass. This implies that the GWPs reported are meaningful only if one wants to compare the potential contribution to the greenhouse effect of equal masses of different gas mixtures.

However, as a matter of fact, the RPC detectors at the LHC are usually operated at constant fractions of gas volumes exchanges. Therefore, it is also useful to compute the carbon dioxide equivalent (CO$_2$e) of one liter of the various gas mixtures considered, namely the amount (in grams) of CO$_2$ that, if injected into the atmosphere, would contribute to the greenhouse effect for the same amount of the one liter of gas mixture considered. This CO$_2$e is computed by calculating the amount (in grams) of the various gaseous components in one liter of the gas mixture, multiplying them for their respective GWP, and summing up. The CO$_2$e values computed in this way are also reported in Table \ref{tab:gasmixutre}, and indeed show that the CO$_2$e for one liter of ECO2 and ECO3 is around 4.5 times lower than the CO$_2$e of one liter of STD mixture. Moreover, ECO2 and ECO3 mixtures feature similar values of CO$_2$e because most of it is due to the emission of SF$_6$, which is quite similar in the two cases.

\begin{table*}
\caption{Percentage composition, in volume, of the gas mixtures used for these tests, their GWP with respect to CO$_2$, and their CO$_2$e, in grams, for one liter of mixture. For the calculations of the GWP and CO$_2e$, the gas densities at STP ($p$ = 1013 hPa, $T$ = 273.15K) of the component gases, reported in the penultimate line of the Table and taken from \cite{NIST}, were used. }
\vspace{0.5cm}
\centering
\label{tab:gasmixutre}
  \begin{tabular}{|p{2.0cm} |p{1.2cm} |p{2.1cm} |p{1cm} |p{1.3cm} |p{1.2cm}  |p{1.1cm}| p{1.6cm}| }
\hline
   & R134a (\%) & HFO-1234ze (\%)& CO$_2$ (\%)& i-C$_4$H$_{10}$ (\%)& SF$_6$ (\%)& GWP & CO$_{2}$e  (g/l)\\ \hline  
STD  & 95.2 &     &        & 4.5 & 0.3 & 1485 & 6824 \\ \hline
ECO2 &      &  35 & 60     & 4   & 1   & 476 & 1522 \\ \hline
ECO3 &      &  25 & 69     & 5   & 1   & 527 & 1519 \\ \hline
Density (g/l) & 4.68 & 5.26 &  1.98 & 2.69  & 6.61  &  & \\ \hline
GWP & 1430  &  7 & 1 & 3 & 22800   &  & \\ \hline
\end{tabular}
\end{table*}

\begin{figure}
\centering 
\includegraphics[width=\linewidth]{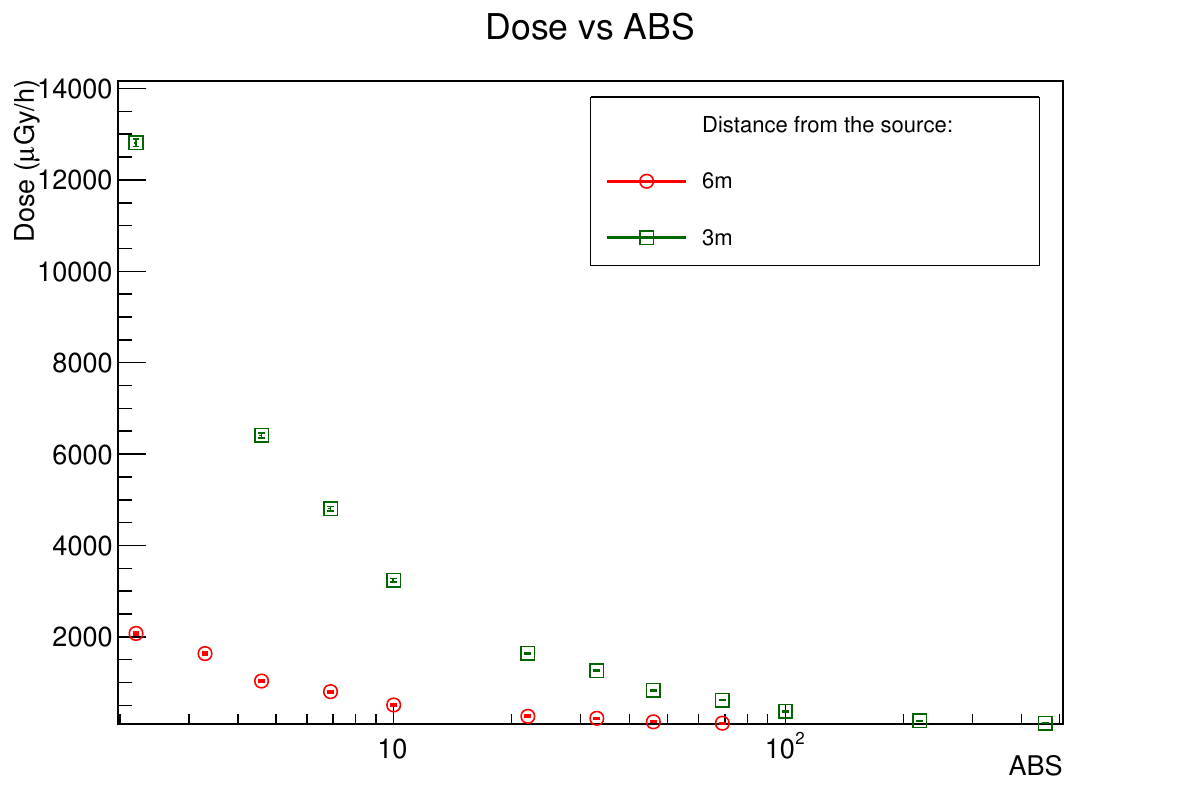}
\caption{Dose measured in the GIF++, at 3 and 6 m from the $^{137}$Cs source, as a function of the ABS}.
\label{fig:Dose}
\end{figure}

Due to the fact that the detectors under test were placed at different distances from the $^{137}$Cs source, different rates of photons were impinging on them, even when the same filter was placed in front of the  source. On the other hand, meaningful comparisons across the various detectors are to be done at the same flux of impinging photons, therefore, using different filter configurations. This strategy could be actuated by using a dosimeter, mod. MIRION RDS-31iTx S/R \cite{dosimeter}, with which the actual gamma ray dose at various positions inside the GIF++ irradiated area was measured. The measured doses at two positions, 3 and 6 m from the source, corresponding to the positions of the trolleys where the detectors were hosted, is shown in Figure~\ref{fig:Dose}, as a function of the ABS used. From the figure, it is clear that an increase in the ABS has the result of reducing the dose measured at a certain position and that, fixed the ABS, the dose measured reduces when moving away from the $^{137}$Cs source, as expected.

\section{Experimental Results}
\label{sec:tests}

Results from the tests described in this paper will be presented in two parts: first, the idea is to assess if the chambers, listed in the previous section, performed, without any irradiation from the $^{137}$Cs source, in similar ways when filled, in sequence, with the STD, ECO2 and ECO3 gas mixtures. For this purpose, in particular, chamber efficiency, current drawn, and cluster size will be examined.

The chambers efficiencies were measured requiring that the coincidence between the signals from the chambers under test and the trigger signal provided by the scintillators was in an acceptance time window of 20 ns. This guaranteed that accidentals, deriving from the random background of photons, constituted a negligible fraction with respect to the total amount of events acquired. 

In order to easily compare data from different chambers, of different areas, here we will graph current density, obtained from the measured values of the absorbed current of the HV power supply channels connected to the chambers under test, and dividing it by the area computed using the dimensions reported in Table \ref{tab:chamberdata}. It is a parameter important for estimating possible aging effects in these detectors.

Also, in order to compare data collected at different environmental conditions, efficiency and current density are plotted here as a function of a variable which we will denote as $HV_{\text{eff}}$. This was computed from the actual voltage $HV_{\text{app}}$ applied to the chambers using the following formula:

\begin{equation}
HV_\text{eff} = HV_{\text{app}} \frac{p_0T}{pT_0}
\end{equation}

\noindent where $p$ is the atmospheric pressure and $T$ the temperature when the measure actually took place, and $p_0$ and $T_0$ are arbitrarily chosen reference values. Here we chose $p_0$ = 990 hPa and $T_0$ = 293.15 K, so that to be close to the average values of temperature and pressure actually measured at GIF++. This procedure is standardly followed in many gaseous detectors studies, and in particular for RPCs \cite{EffectiveVoltage}.

The second part of the tests was intended to study the performance of the chambers when irradiated with the photons from the $^{137}$Cs source. Multiple ABS factors were used, causing hit rates in the RPCs up to several hundred Hz/cm$^2$, and the measurements were repeated with the STD, ECO2 and ECO3 mixtures. For the sake of brevity, here we will report results obtained with just few (up to five) of the ABS settings actually used.

Note also that, often, we will not show the plots obtained from all the chambers under test, but just part of them. Unless otherwise specified, they are intended to be representative of the general behaviour of the chambers.

\subsection{Chambers performance with no irradiation}
\label{subsec:noirr}

\begin{figure*}
\centering 
\includegraphics[width=0.48\textwidth]{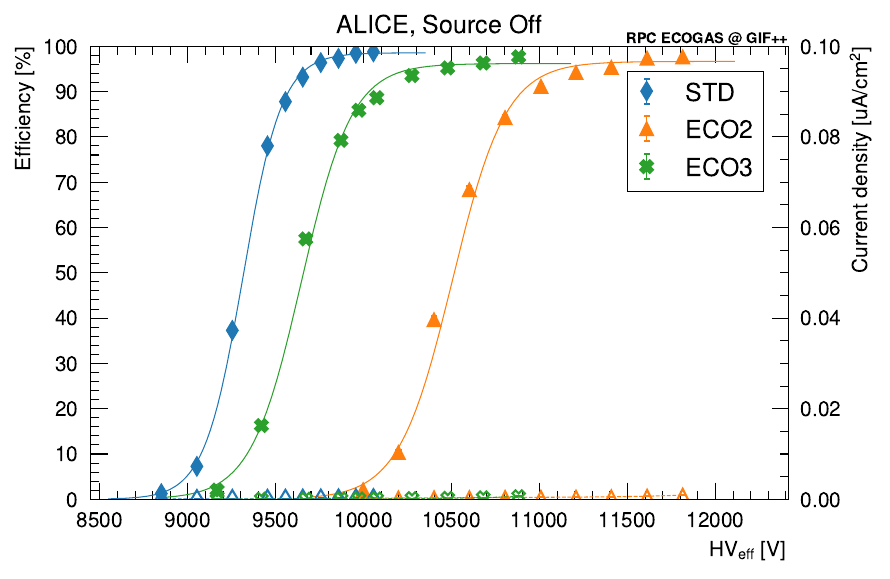}
\includegraphics[width=0.48\textwidth]{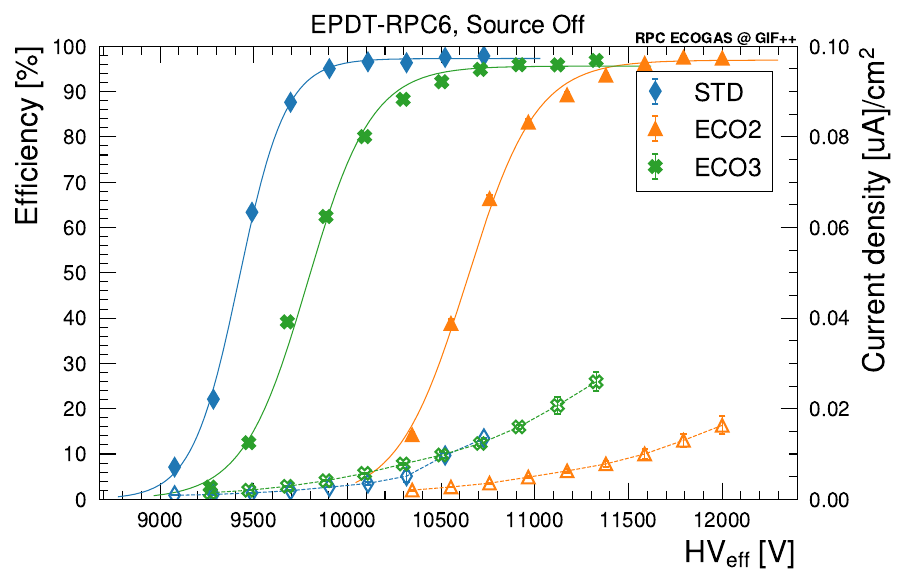} \\
\includegraphics[width=0.48\textwidth]{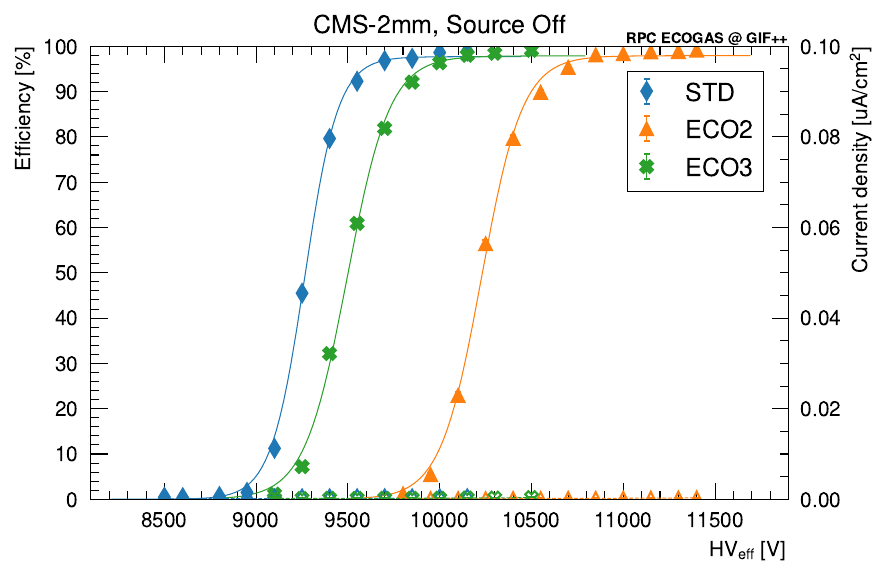}
\includegraphics[width=0.48\textwidth]{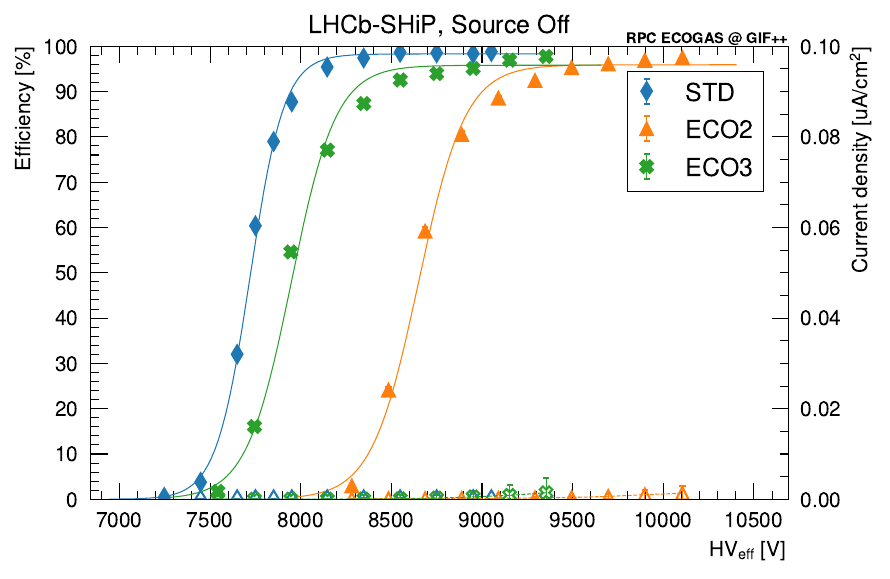}
\caption{Top Left: Efficiency and current density, as a function of HV$_{eff}$, measured for the ALICE chamber without irradiation, filled, in sequence, with the STD, ECO2 and ECO3 gas mixtures. Top Right: Same, but for the CERN EP-DT chamber; Bottom Left: Same, but for the CMS chamber; Bottom Right: Same, but for the LHCb/SHiP chamber.}
\label{fig:EFF1}
\end{figure*}

The efficiency and current density curves, measured without any irradiation, for the ALICE, CERN EP-DT, CMS and LHCb/SHiP detectors, filled, in sequence, with the STD, ECO2 and ECO3 gas mixtures, are shown as a function of $HV_{\text{eff}}$ in the panels of Figure~\ref{fig:EFF1}. As it can be seen, all efficiency curves reach a plateau efficiency above 95\%. The CMS chamber reaches higher efficiencies at lower voltage with respect to the chambers from ALICE and CERN EP-DT, due to the fact that it is a double-gap chamber, and signals from the two gaps add up on the same readout strips. This is also at the base of the fact that the transition from low to high efficiency, for the CMS chamber, is generally steeper with respect to the other devices. The chamber from LHCb/SHiP features the efficiency curves positioned at the lowest voltages, due to the fact that its gap is just 1.6 mm thick, while the others are 2 mm. In all cases, the efficiency curve for the ECO2 mixture is the rightmost one, featuring full efficiency above 11.5 kV for 2 mm gap RPCs.

Note that the current density measured with the CERN EP-DT chamber, is larger than the one measured with the other chambers; this is probably related to construction details and to the fact that this particular chamber was also used for other tests in the past.

 For an easy comparison across the results obtained from the various chambers, the efficiency curves were fitted with a logistic function, of the form:
\begin{equation}
\mathcal{E}(HV_{\text{eff}}) = \frac{\mathcal{E}_{\text{max}}} { 1+ e^{ - \beta (HV_{\text{eff}} - HV_{50} )   }  }       
\end{equation}

 Fit parameters were the plateau efficiency $\mathcal{E}_{\text{max}}$ (namely the asymptotic value of the logistic function fitted), the voltage at which the chamber efficiency reaches 50\% of $\mathcal{E}_{\text{max}}$, denoted here as HV$_{50}$,  and the steepness of the curve, $\beta$. From these parameters, we also extracted another parameter, indicated here as $HV_{\text{knee}}$, which is the voltage at which the chamber efficiency reaches 95\% of the plateau efficiency $\mathcal{E}_{\text{max}}$. $HV_{\text{knee}}$, also briefly indicated as the "knee" of the efficiency curves, is a good indication of the optimal voltage for operating the chamber considered, and is given by the expression:

\begin{equation}
HV_{\text{knee}} = HV_{50} + \frac{1}{\beta} \ln{ \frac{0.95}{0.05} }     
\end{equation}

 The values of all these parameters, obtained from the fit, for the detectors from ALICE, CMS, CERN EP-DT and LHCb/SHiP, and the three gas mixtures used, are reported in Table \ref{tab:FitParameters}. In the same Table \ref{tab:FitParameters} also the current densities $J_\text{knee}$, as previously defined, measured at $HV_{\text{knee}}$, with the various chambers, are listed. 
 Results from the ATLAS chamber are not present in Figure~\ref{fig:EFF1} and Table \ref{tab:FitParameters}, since this chamber was not used for these tests, but for charge measurements, as will be discussed later on in this section. 

\begin{table*}
\caption{Parameters obtained from the fit with the logistic function discussed in the text, for the chambers under test without irradiation. Also $HV_\text{knee}$ and current density $J_\text{knee}$ at $HV_\text{knee}$ are listed in the table. Errors on the $HV$ values are indicated implicitly, while $J_\text{knee}$ behaviour is discussed in the following section.}
\vspace{0.5cm}
\centering
\label{tab:FitParameters}
  \begin{tabular}{|p{1.8cm} |p{1cm} |p{2.1cm}  |p{2.1cm} |p{1.7cm} |p{1.8cm} |p{2cm} |}
\hline
 Detector  & Gas Mix. & $HV_{50}$ (kV) & $HV_\text{knee}$ (kV) & $\mathcal{E}_\text{max}$ ($\%$) & $\beta$ (kV$^ {-1}$) & $J_\text{knee}$  (nA/cm$^2$)\\ \hline  
ALICE  & STD & 9.31 & 9.64  & 98.6$\pm$0.1& 9.1 $\pm$0.1 & 0.148$\pm$0.001\\ \hline
ALICE  & ECO2 & 10.50 & 10.91 & 96.7$\pm$0.2 & 6.9 $\pm$0.1& 0.35 $\pm$ 0.01\\ \hline
ALICE  & ECO3 & 9.64 & 10.07 & 96.2$\pm$0.2 & 6.9 $\pm$0.1 & 0.10 $\pm$ 0.01\\ \hline\hline
CMS  & STD & 9.27 & 9.69 & 97.7$\pm$0.6 & 11.0 $\pm$0.4 & 0.2$\pm$0.1\\ \hline
CMS  & ECO2 & 10.23 & 10.71 & 97.9$\pm$0.7 & 9.0 $\pm$0.4& 0.2$\pm$0.1\\ \hline
CMS  & ECO3 & 9.49 & 9.98 & 97.9$\pm$0.9 & 8.6 $\pm$0.5& 0.2$\pm$0.1\\ \hline\hline
EP-DT  & STD & 9.42 & 9.79 & 97.3$\pm$0.1 & 8.0 $\pm$0.1   & 2.25$\pm$0.04\\ \hline
EP-DT  & ECO2 & 10.64 & 11.18 & 96.9$\pm$0.2 & 5.4$ \pm$0.1 & 6.3$\pm$0.4\\ \hline
EP-DT  & ECO3 & 9.79 & 10.29 & 95.6$\pm$0.2& 5.8$ \pm$0.1 & 7.7$\pm$0.1\\ \hline\hline
LHCb/SHiP & STD & 7.72 & 8.00 & 98.3$\pm$0.1 & 10.7 $\pm$0.1 & 0\\ \hline
LHCb/SHiP  & ECO2 & 8.65 & 9.03 & 95.9$\pm$0.1 & 7.7$ \pm$0.1& 0.21$\pm$0.08\\ \hline
LHCb/SHiP  & ECO3 & 7.95 & 8.31 & 95.8$\pm$0.1 & 8.1$ \pm$0.1& 0.1$\pm$0.1\\ \hline

\end{tabular}
\end{table*}

\begin{figure*}
\centering 
\includegraphics[width=0.32\textwidth]{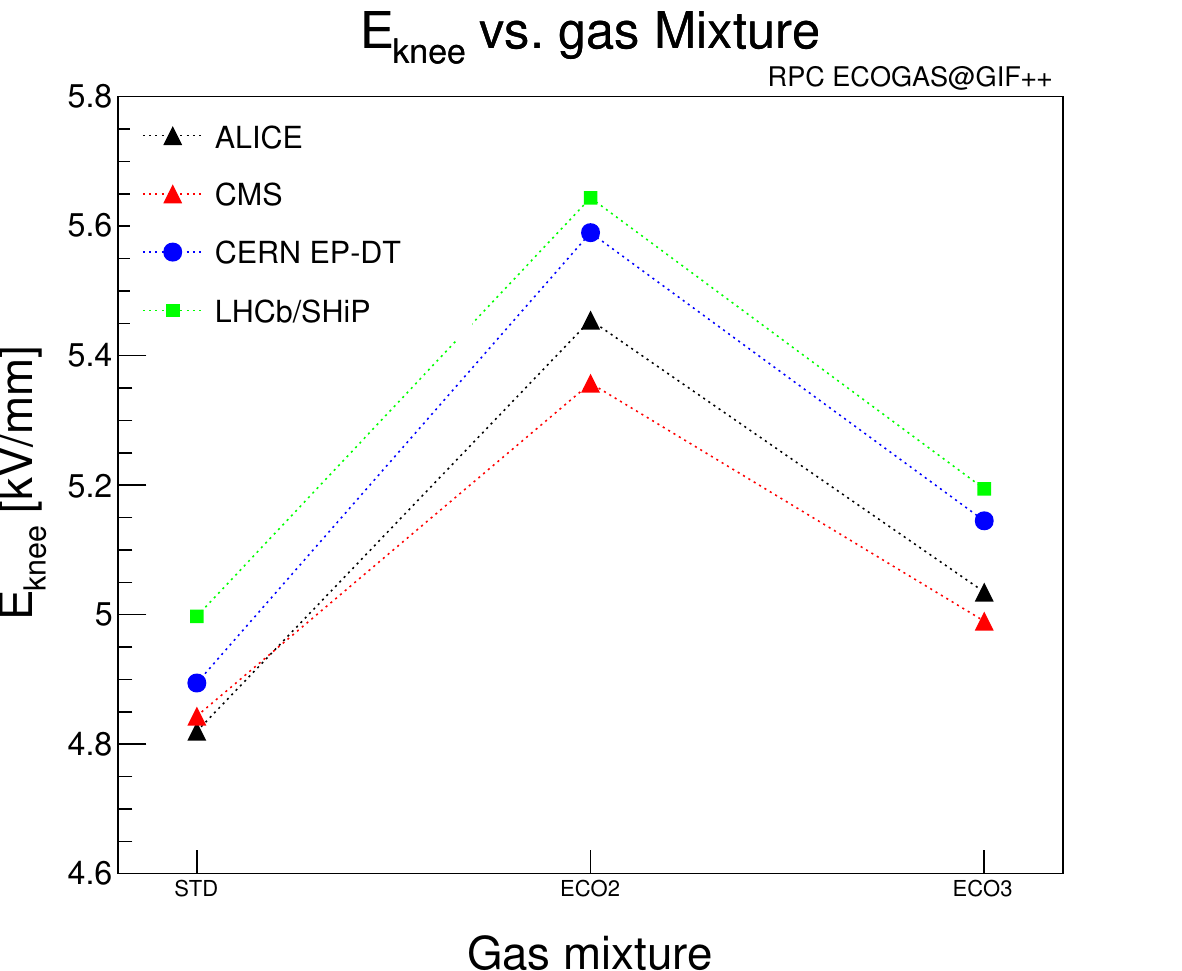}
\includegraphics[width=0.32\textwidth]{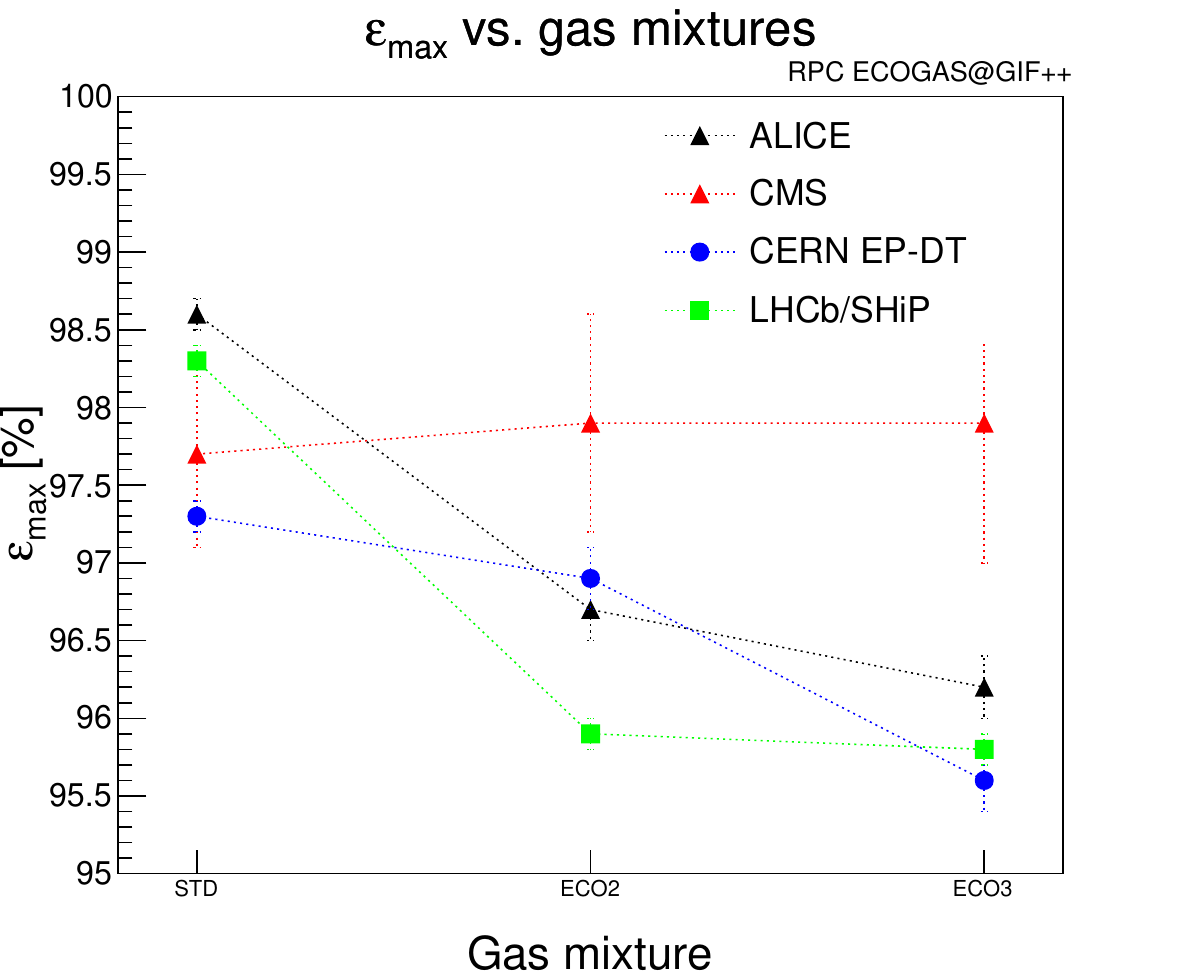}
\includegraphics[width=0.32\textwidth]{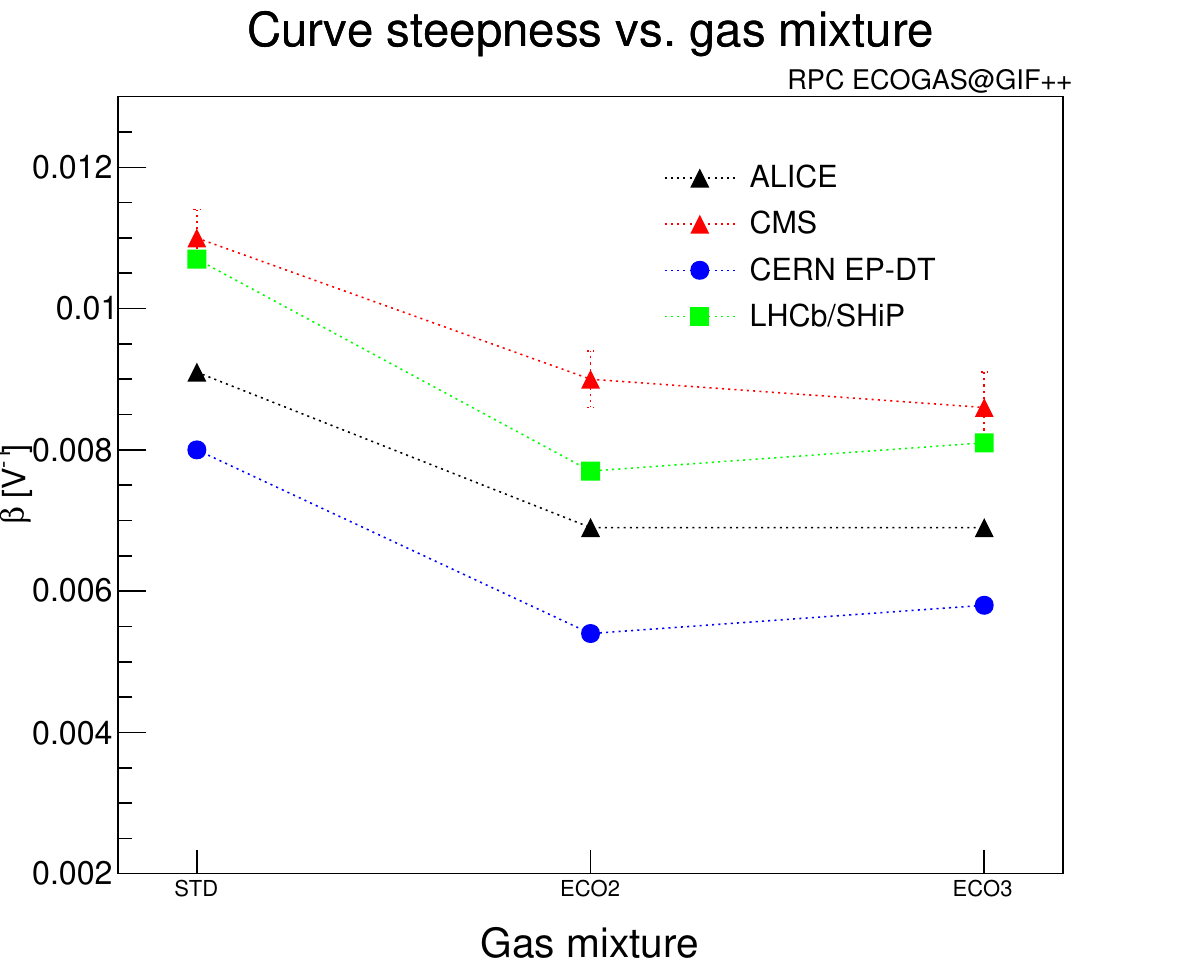}
\caption{Left: Electric field at the knee $E_{knee}$ for the ALICE, CERN EP-DT, CMS and LHCb/SHiP chambers, when filled with STD, ECO2 and ECO3 gas mixtures; Center: plateau efficiency $\mathcal{E}_\text{max}$, for the same chambers and gas mixtures; Right: steepness $\beta$ of the logistic functions for the same chambers and gas mixtures.}
\label{fig:PlotFromTable4}
\end{figure*}

Using the data reported in Table \ref{tab:FitParameters} we can easily compare the performance of the various chambers when filled with the different gas mixtures examined. For comparison across chambers characterized by a different gas gap thickness $d$, it is useful to refer to the electric field in the gas gap at the knee $E_\text{knee}$ = $HV_\text{knee}$/$d$. With respect to the STD mixture, the ECO3 is characterized by a value of $E_\text{knee}$ between 200 and 280 V/mm higher, while the ECO2 mixture by a value of $E_\text{knee}$ between 650 and 700 V/mm higher. The increase of $E_\text{knee}$ from STD to ECO2 is related to the replacement of TFE with HFO-1234ze, characterized, as already pointed out, by a lower first effective Townsend coefficient for the same electric field strength. The decrease of $E_\text{knee}$ from ECO2 to ECO3 is due to the 10\% larger percentage of HFO-1234ze in ECO2. $E_\text{knee}$  values for the ALICE, CERN EP-DT, CMS and LHCb/ShiP chambers are shown in the left panel of Figure~\ref{fig:PlotFromTable4}, where both these effects are clearly visible. It also shows that $E_{knee}$ for the LHCb/SHiP chamber is the largest in this set, because of the reduced gas thickness available for amplification processes. The double gap layout is at the base of the lowest $E_\text{knee}$ for the CMS chamber.

From the data shown in Table \ref{tab:FitParameters}, it can also be noted that, as a general trend, the plateau efficiency $\mathcal{E}_\text{max}$ decreases when passing from STD, to ECO2, and to ECO3 mixtures, namely when the fraction of CO$_{2}$ in the gas mixture considered increases. In the last two cases, this effect might be related to the reduced number of primary ion-electron pairs produced by the impinging particles, as well as the increased distance among them. In fact, it is more evident for the RPC characterized by the thinnest gas gap here, from LHCb/SHiP, while is virtually not present for the CMS detector, because of the double gap layout and the consequent larger total thickness of the gas layers. $\mathcal{E}_\text{max}$ is shown in the central panel of Figure~\ref{fig:PlotFromTable4}, for the four chambers under test and the three gas mixtures considered.

The steepness $\beta$ of the logistic functions fitted to the efficiency curves is shown in the right panel of Figure~\ref{fig:PlotFromTable4}. As expected, steepness is larger for the chamber from CMS, because of the double gap configuration, and for the LHCb/ShiP chamber, because of the thinner gas gap, signalling in both cases a fast transition from zero to full efficiency. Moreover, the steepness of the efficiency curves of all the four chambers under test decreases, when using ECO2 and ECO3, with respect to the STD mixture, showing that the transition from low to high efficiency in these cases takes place more slowly. This is a hint of the fact that charge distributions for ECO2 and ECO3 are wider with respect to STD mixture, as it was experimentally verified and discussed in the following.

\begin{figure*}
\centering 
\includegraphics[width=0.98\textwidth]{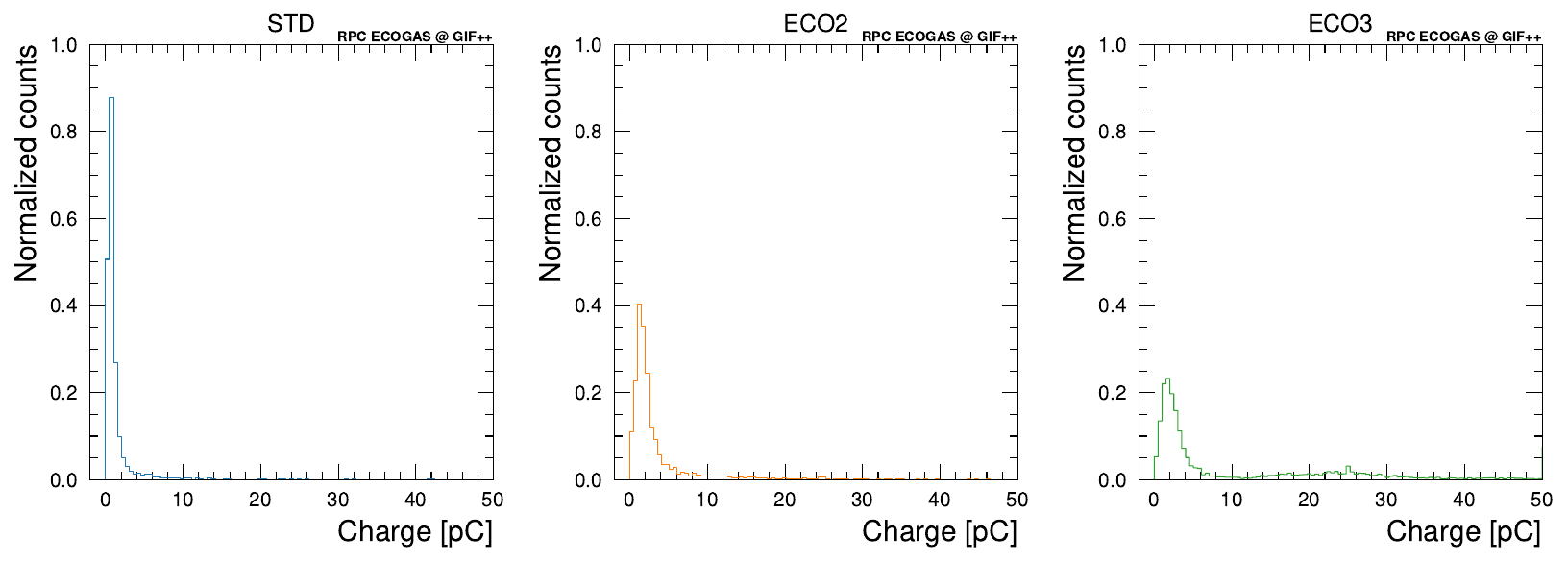}
\caption{Charge distributions of the signals induced on the 3 cm wide strip of the ATLAS RPC, measured at $HV_\text{knee}$, for the STD, ECO2 and ECO3 gas mixtures, from left to right, respectively.}
\label{fig:AtlasChargeOff}
\end{figure*}

Charge distributions of the induced signals at $HV_\text{knee}$ were measured with the ATLAS chamber, using the digitizer it was equipped with, for the three gas mixtures tested, and are shown in Figure~\ref{fig:AtlasChargeOff}. In this case, a "finger" scintillator was put in front of the one readout strip, in order to select only events actually passing through it. A peak at the left end of the distributions is clearly visible, with an average charge of few pC, corresponding to the events where a Townsend avalanche developed in the gas gap, which is the case for most of them. Moreover, the horizontal scale is suitably calibrated in order to show also the presence, for ECO2 and ECO3, of a small fraction of events with a charge typically an order of magnitude larger than the one corresponding to avalanches in the same mixture. These events, however, are characterized by an average charge lower then streamers and are not delayed in time from a precursor as streamers usually are. Therefore they cannot be classified as streamers and fall in a somewhat intermediate category, whose properties need to be further investigated.

Basically, these results show that with respect to the STD gas mixture, the increased percentage of CO$_2$ in the eco-friendly mixtures broadens the charge distributions and increases the fraction of signals at a larger charge. Vice versa, the fraction of these events is smaller for larger HFO-1234ze concentrations because the quenching properties of HFO-1234ze limit the transition probability from avalanches to streamers.

\begin{figure}[!h]
\centering 
\includegraphics[width=0.45\textwidth]{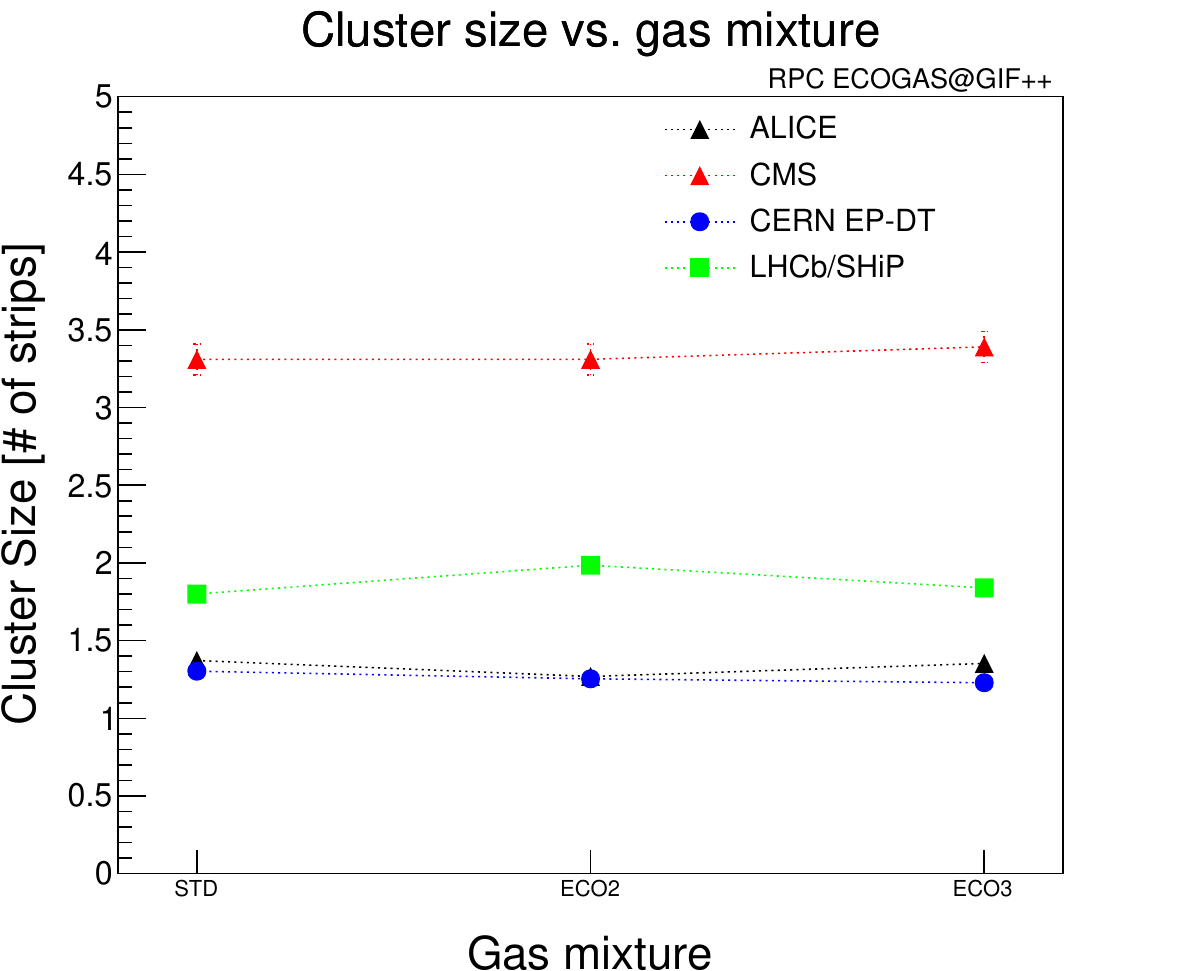}
\caption{Cluster size for the ALICE, CERN EP-DT, CMS and LHCb/SHiP chambers, expressed in number of strips, when filled with STD, ECO2 and ECO3 gas mixtures.}
\label{fig:ClusterSize}
\end{figure}

The average cluster size, intended as the average number of adjacent strips transmitting an over-threshold signal related to the passage of the same muon, is shown in Figure~\ref{fig:ClusterSize} for the detectors from ALICE, CMS, CERN EP-DT and LHCb/SHiP, and the three gas mixtures used. Values reported in the plot were measured at $HV_\text{knee}$ and are expressed in number of strips, this implying that they actually refer to different cluster physical dimensions. Moreover, the measured values depend on the amplifications and electronic thresholds used. Nevertheless, the cluster size stays approximately constant when passing from STD to ECO2 and ECO3 gas mixtures, indicating that spatial resolution is not significantly affected.

A similar behaviour has been verified for what concerns time resolution as well, which stays basically constant in the 1$\div$2 ns range. Nevertheless, the technical layout used for the tests discussed in this paper does not allow a direct precise measure of such a parameter, which will be accurately measured in the ongoing tests.

\subsection{Chambers performance when irradiated from the $^{137}$Cs source}

\begin{figure*}
\centering 
\includegraphics[width=0.32\textwidth]{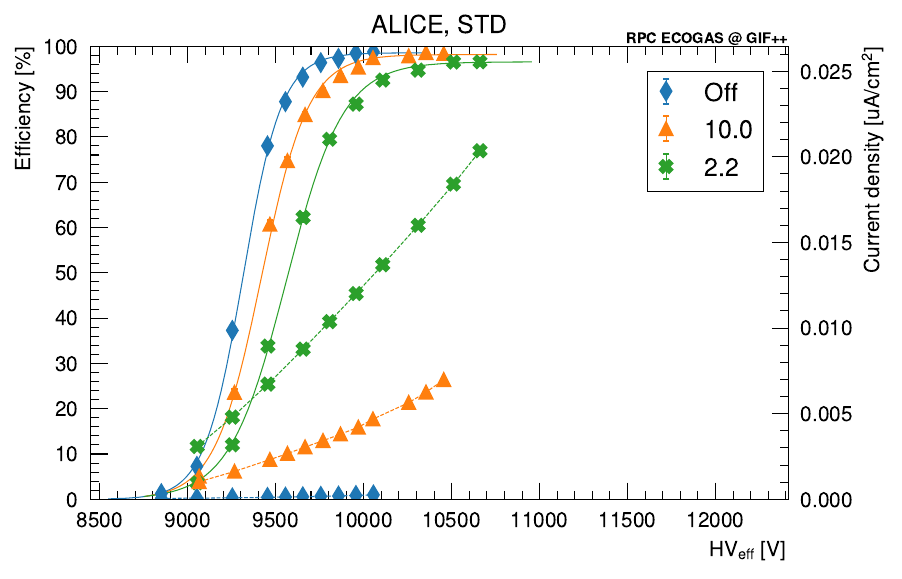}
\includegraphics[width=0.32\textwidth]{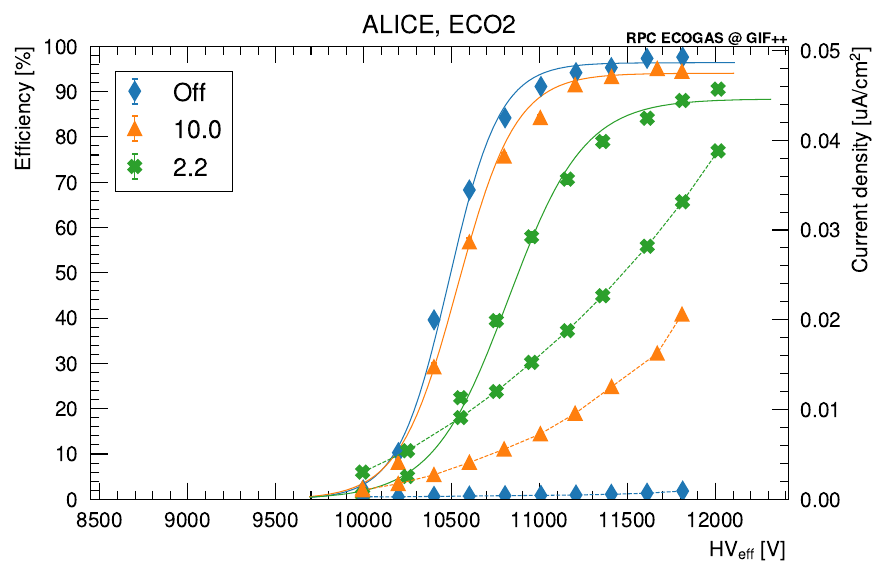}
\includegraphics[width=0.32\textwidth]{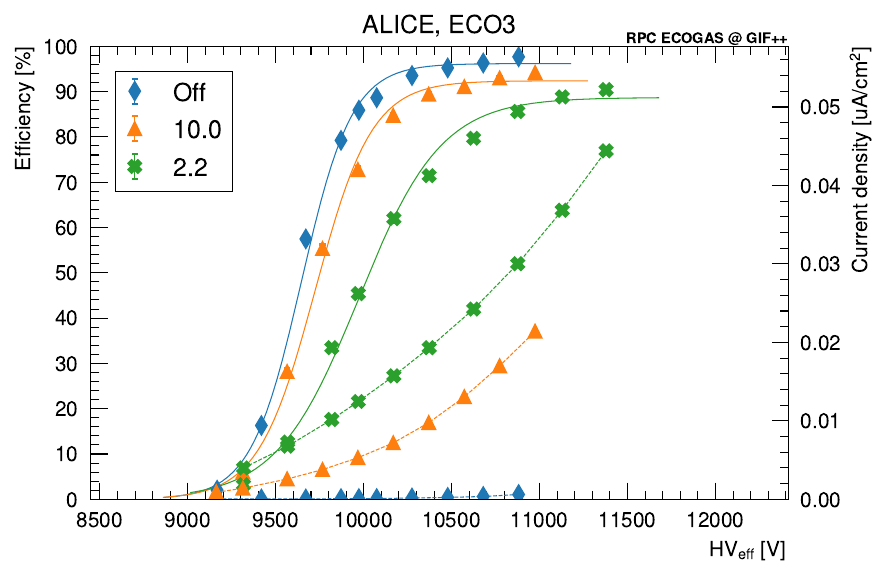}
\caption{Left: Efficiency and current density as a function of $HV_\text{eff}$, for the ALICE chamber filled with the STD mixture, obtained with no irradiation, and with ABS= 10 and 22; Center: Same, but with the same chamber filled with the ECO2 mixture; Right: Same, but with the same chamber filled with the ECO3 mixture.}
\label{fig:AliceHighRate}
\end{figure*}

Efficiency and current density curves as a function of $HV_\text{eff}$, obtained with the ALICE chamber, when subjected to irradiation from the $^{137}$Cs source, are plotted in Figure~\ref{fig:AliceHighRate}, and in particular, the left panel refers to the chamber filled with the STD gas mixture, central and right panels to the chamber filled with ECO2 and ECO3, respectively. Results reported here were taken using the absorption filters ABS = 10 and ABS = 2.2, which correspond to an absorbed dose on the chamber of 510 and 2070 $\mu$Gy/h, as can be deduced from Figure~\ref{fig:Dose} considering that the ALICE chambers was on the trolley at 6 m from the source. On the same plots, also results obtained without any irradiation are superimposed, for ease of comparison.

\begin{figure*}
\centering 
\includegraphics[width=0.32\textwidth]{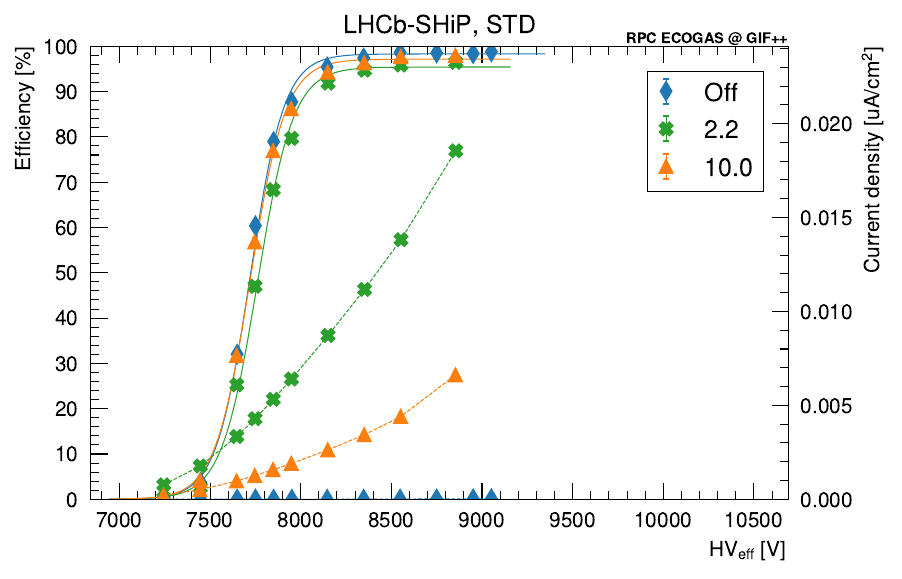}
\includegraphics[width=0.32\textwidth]{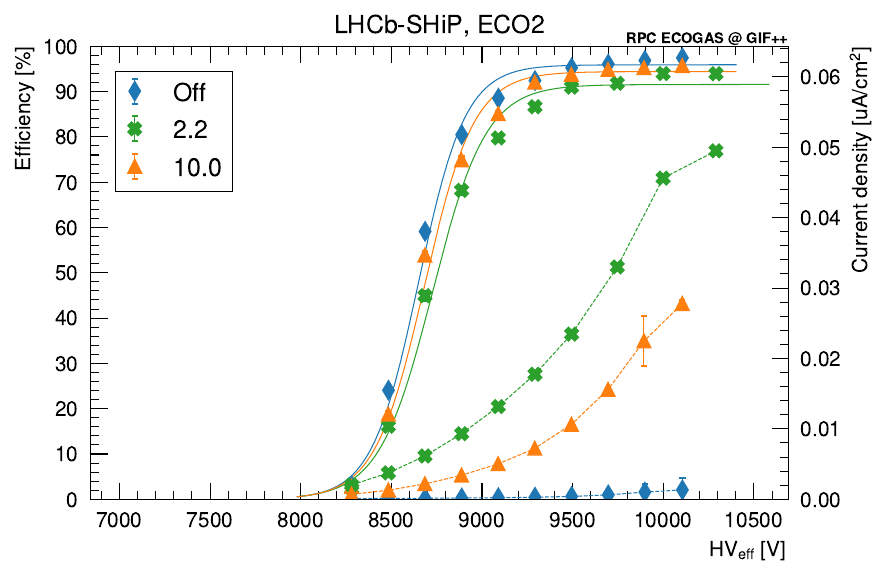}
\includegraphics[width=0.32\textwidth]{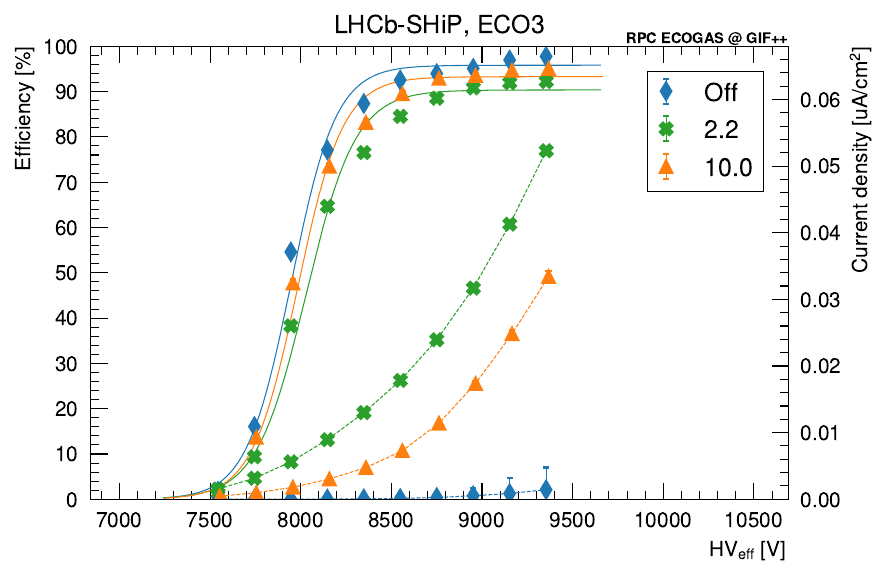}
\caption{Left: Efficiency and current density as a function of $HV_\text{eff}$, for the LHCb/SHiP chamber filled with the STD mixture, obtained with no irradiation, and with ABS= 10 and 22; Center: Same, but with the same chamber filled with the ECO2 mixture; Right: Same, but with the same chamber filled with the ECO3 mixture.}
\label{fig:LHCbHighRate}
\end{figure*}

Similar plots were also obtained with the chambers from CMS, CERN EP-DT and LHCb/SHiP; for exemplification, only the ones using data collected with the LHCb/SHiP chamber are shown in Figure~\ref{fig:LHCbHighRate}. Moreover, the efficiency data points resulting from these tests were fitted with a logistic function of the type described in Eq.(2), and the relative fit parameters extracted, analogously to what done and described in Section \ref{subsec:noirr}.

\begin{figure*}
\centering 
\includegraphics[width=0.32\textwidth]{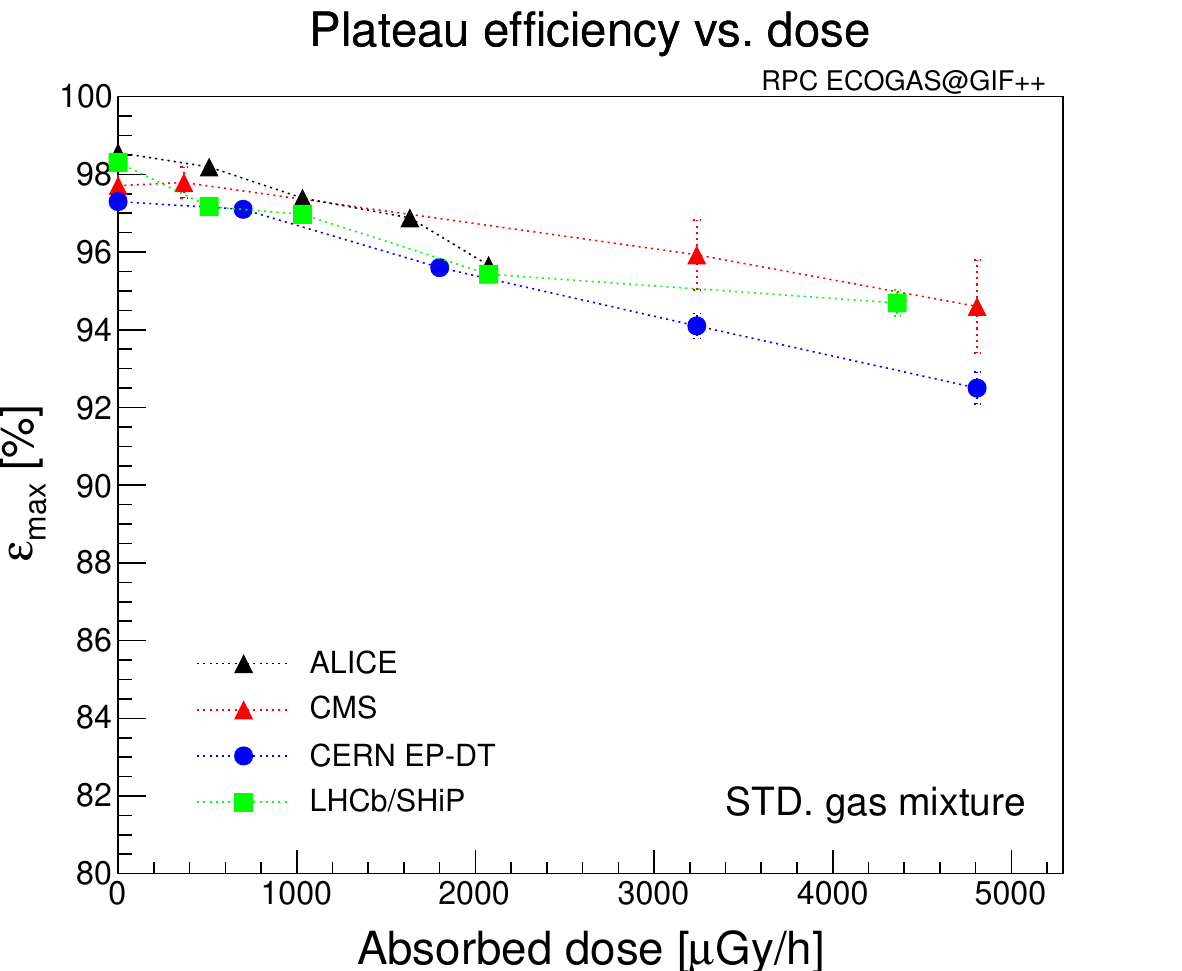}
\includegraphics[width=0.32\textwidth]{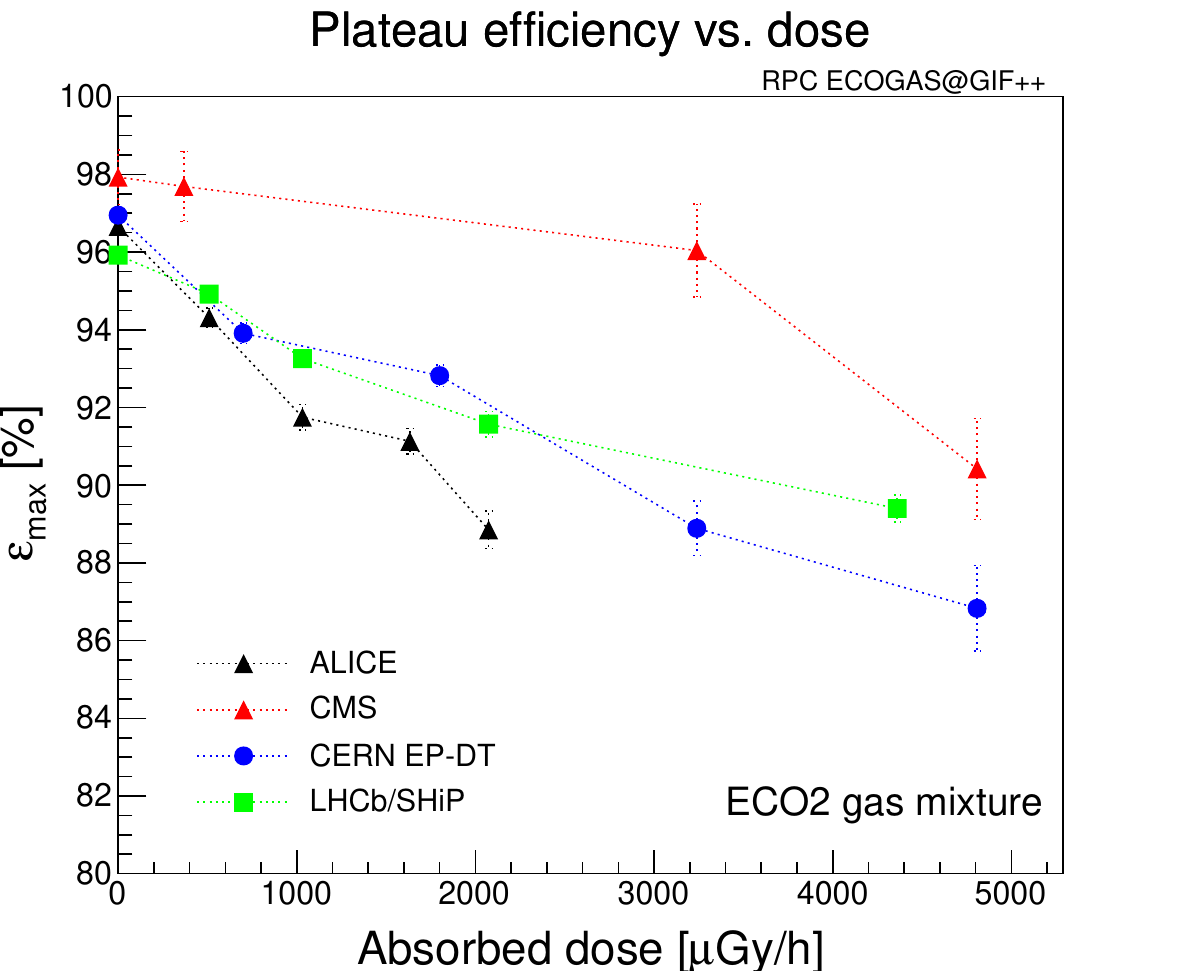}
\includegraphics[width=0.32\textwidth]{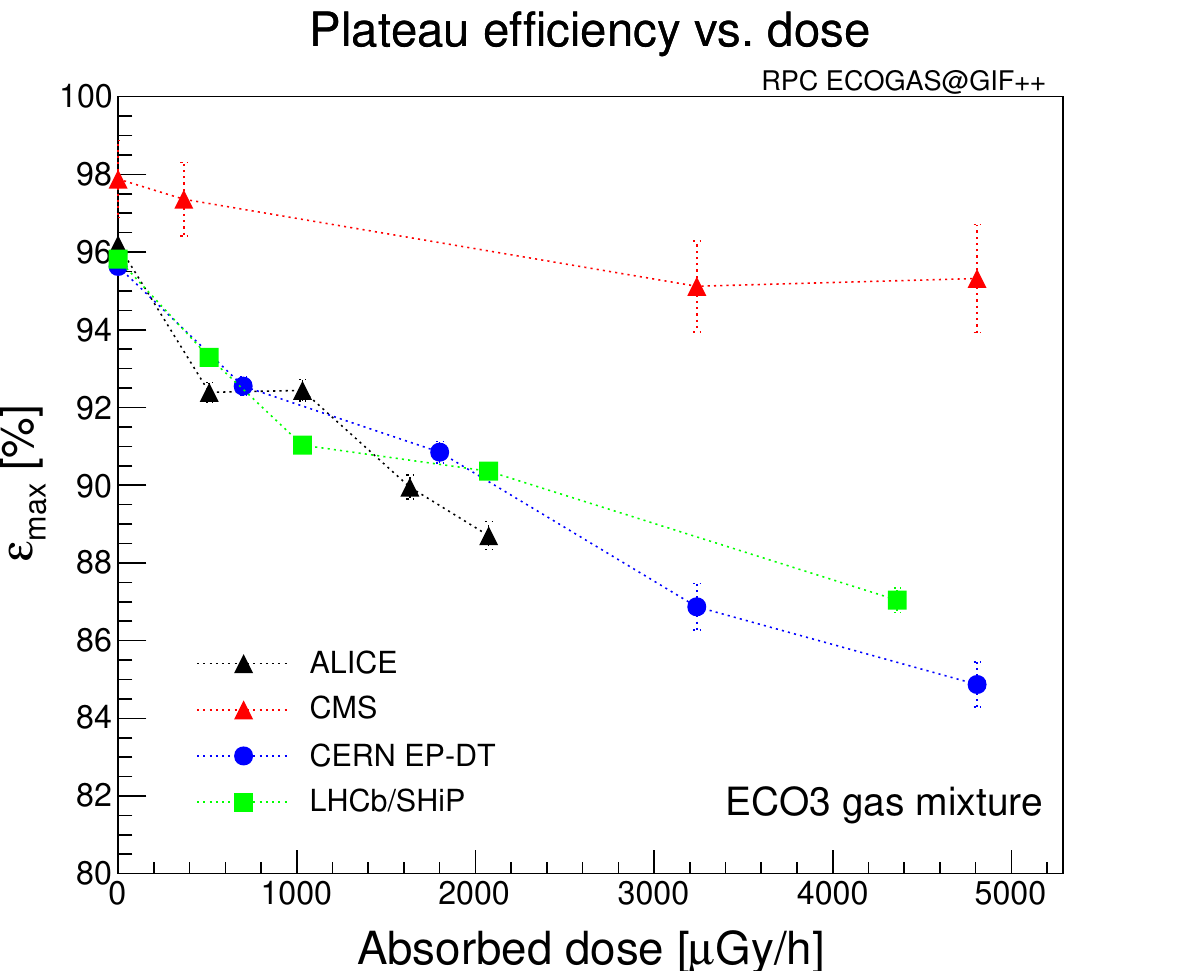}
\caption{Left: $\mathcal{E}_{\text{max}}$ as a function of the absorbed dose for the four chambers under test when filled with the STD mixture; Center: Same, but with the same chambers filled with the ECO2 mixture; Right: Same, but with the same chambers filled with the ECO3 mixture.}
\label{fig:MaxEffDose}
\end{figure*}

There are three basic aspects that can be inferred from the results reported in Figures ~\ref{fig:AliceHighRate} and ~\ref{fig:LHCbHighRate}, and the others from CMS and CERN EP-DT groups. The first is that, as a general trend, the plateau efficiency $\mathcal{E}_{\text{max}}$ decreases when increasing the absorbed dose. In addition, this decrease appears to be larger when using ECO2 and ECO3 mixtures with respect to STD. $\mathcal{E}_{\text{max}}$, as obtained from the fit with the logistic function, versus the absorbed dose is shown for the four chambers under test in the three panels of Figure~\ref{fig:MaxEffDose}, and indeed shows this behaviour, which can be quantified with a 3$\div$4~\% drop in efficiency when passing from 0 to 5000 $\mu$Gy/h using the STD mixture. This has to be compared to an around 10 \%  efficiency drop in the same conditions but using ECO2 and ECO3. Again the efficiency drop for the CMS chamber is relatively contained with respect to the others, due to the double gap configuration. Note that the ALICE chamber was tested only up to around 2000 $\mu$Gy/h, given the lower maximum rate expected at the experimental site with respect to the other experiments.

\begin{figure*}
\centering 
\includegraphics[width=0.32\textwidth]{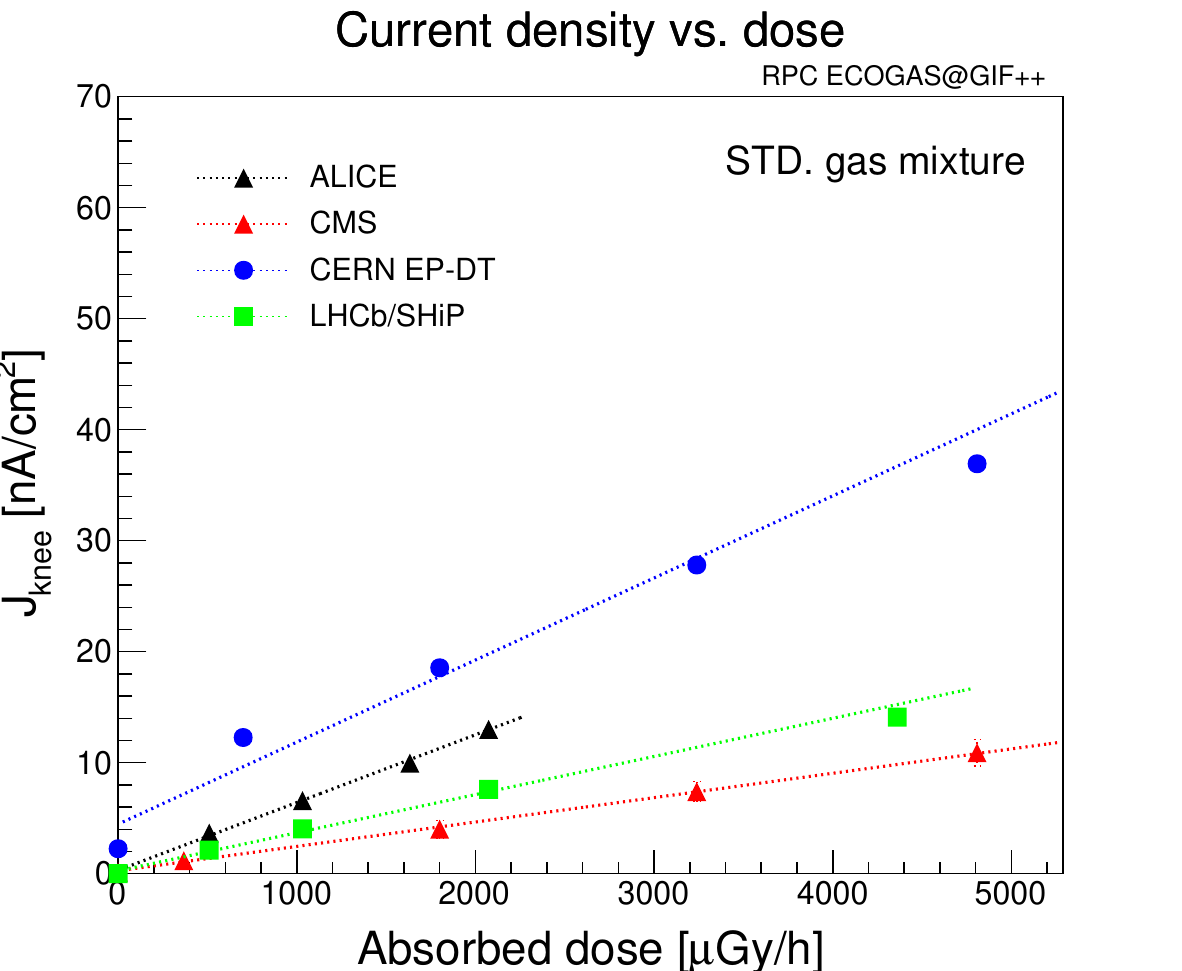}
\includegraphics[width=0.32\textwidth]{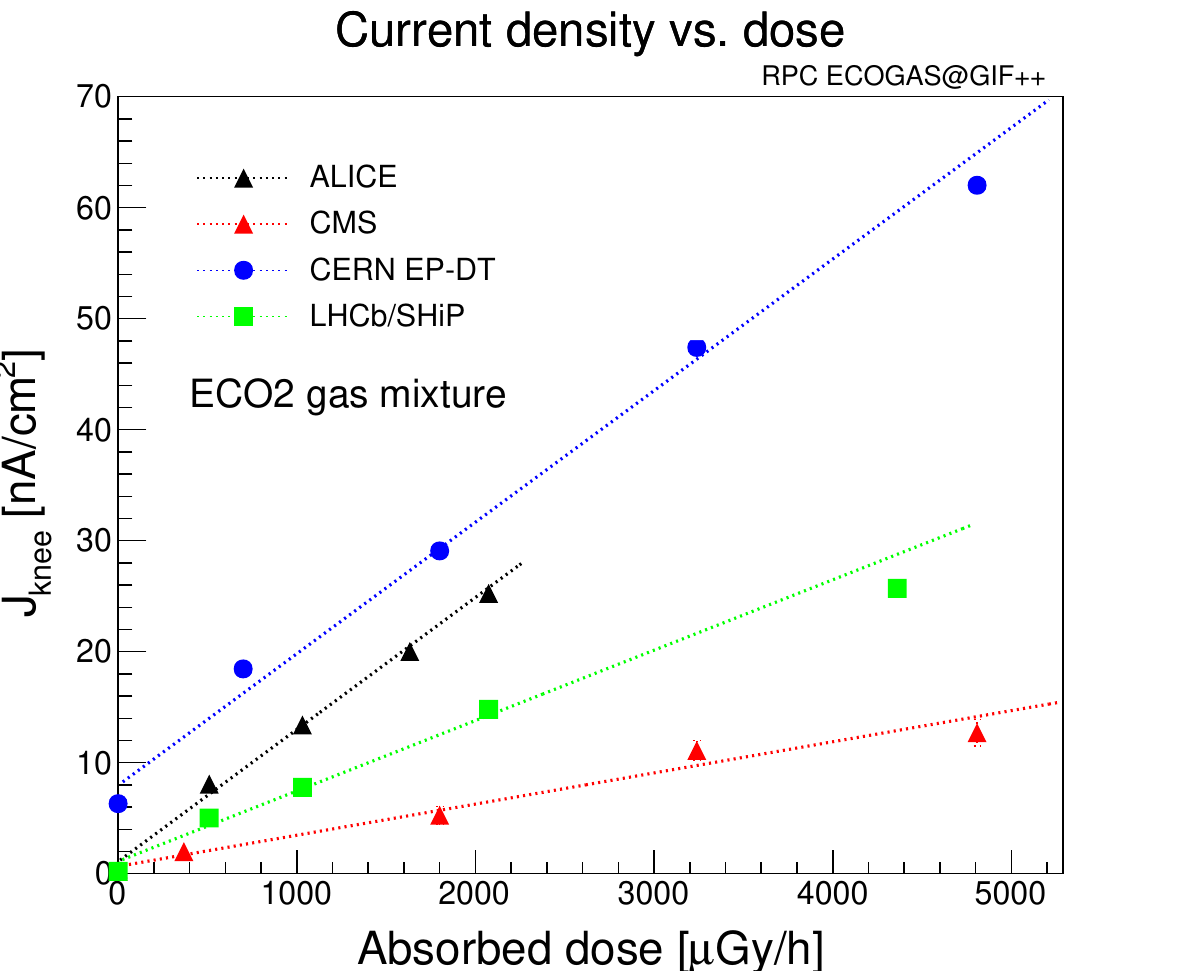}
\includegraphics[width=0.32\textwidth]{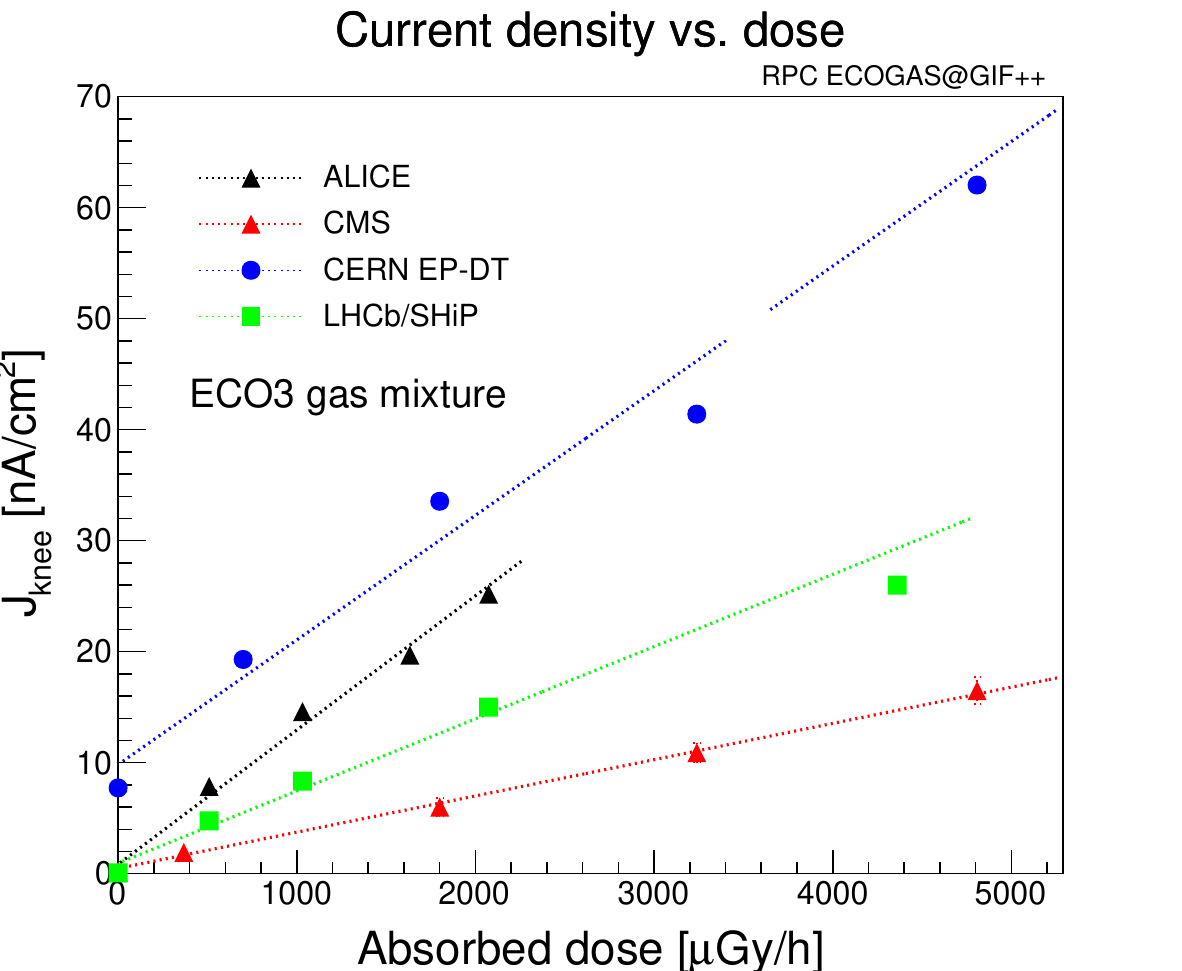}
\caption{Left: Current density $J_\text{knee}$, measured at $HV_\text{knee}$, as a function of the absorbed dose for the four chambers under test when filled with the STD mixture; Center: Same, but with the same chambers filled with the ECO2 mixture; Right: Same, but with the same chambers filled with the ECO3 mixture.}
\label{fig:CurrDensDose}
\end{figure*}

The second aspect is that current densities $J_\text{knee}$, measured at $HV_\text{knee}$, are generally larger when using ECO2 and ECO3 with respect to STD mixture. $J_\text{knee}$ as a function of the absorbed dose is shown in Figure~\ref{fig:CurrDensDose} for the four chambers under test and the three gas mixtures used. Since one could expect an approximately linear proportionality between current density and dose absorbed, a fit with a straight line has been superimposed to the data points, demonstrating this proportionality. Overall, the increase in $J_\text{knee}$ when using ECO2 or ECO3 with respect to the STD gas mixture can be quantified in around a factor of two for the single gap chambers; for the CMS chamber this is much reduced, very likely due to the double-gap configuration. Note that the value of $J_\text{knee}$ reported here for the CMS chamber is the average across the ones measured on the relative HV channels. The fact that, even without irradiation, current density for the CERN EP-DT chamber is not negligible has already been noted and commented in the previous Section. Finally, note that the different slopes in the lines plotted in Figure~\ref{fig:CurrDensDose} denote that, as a matter of fact, the various chambers are operating with a different value of the average charge per count. This is related to the different configurations and front-end electronics used.

The increased $J_\text{knee}$ observed can be related to the wider charge distributions shown in Figure~\ref{fig:AtlasChargeOff} for ECO2 and ECO3, the presence of events characterized by a higher charge and, at the end, to the increased fraction of CO$_2$ in ECO2 and ECO3. 

In general, large currents are known to cause aging effects in RPCs, and should be avoided or, at least, limited. One possible solution could be, in principle, adding some iso-butane to the mixture used. Being an hydrocarbon with many vibrational and rotational degrees of freedom, iso-butane is known for its UV photon absorption capability and quenching properties and, in fact, is presently used in the gas mixtures for the RPCs of the LHC experiments exactly for this purpose. However, one has to be cautious about the maximum fraction of iso-butane usable, because the resulting mixture should be not flammable to be used at the LHC experiments, due to the flammability security limits. The exact fraction usable should be measured, with dedicated tests, since this depends not only on the percentage of iso-butane, but also on the other gases in the mixture and their relative fractions.

An increase in the fraction of SF$_6$ would also probably help, for analogous reasons, in particular for its high attachment coefficient for the electrons in the Townsend avalanches, but, as already pointed out and reported in Table \ref{tab:gasmixutre}, this gas is characterized by a very high GWP, which would compromise the eco-friendly characteristics of these mixtures and, moreover, would push at too large operating voltages. These lines of reasoning is at the base of the gas mixtures composition optimization which led to the definition of the ECO2 and ECO3 mixtures tested. 

\begin{figure*}
\centering 
\includegraphics[width=0.32\textwidth]{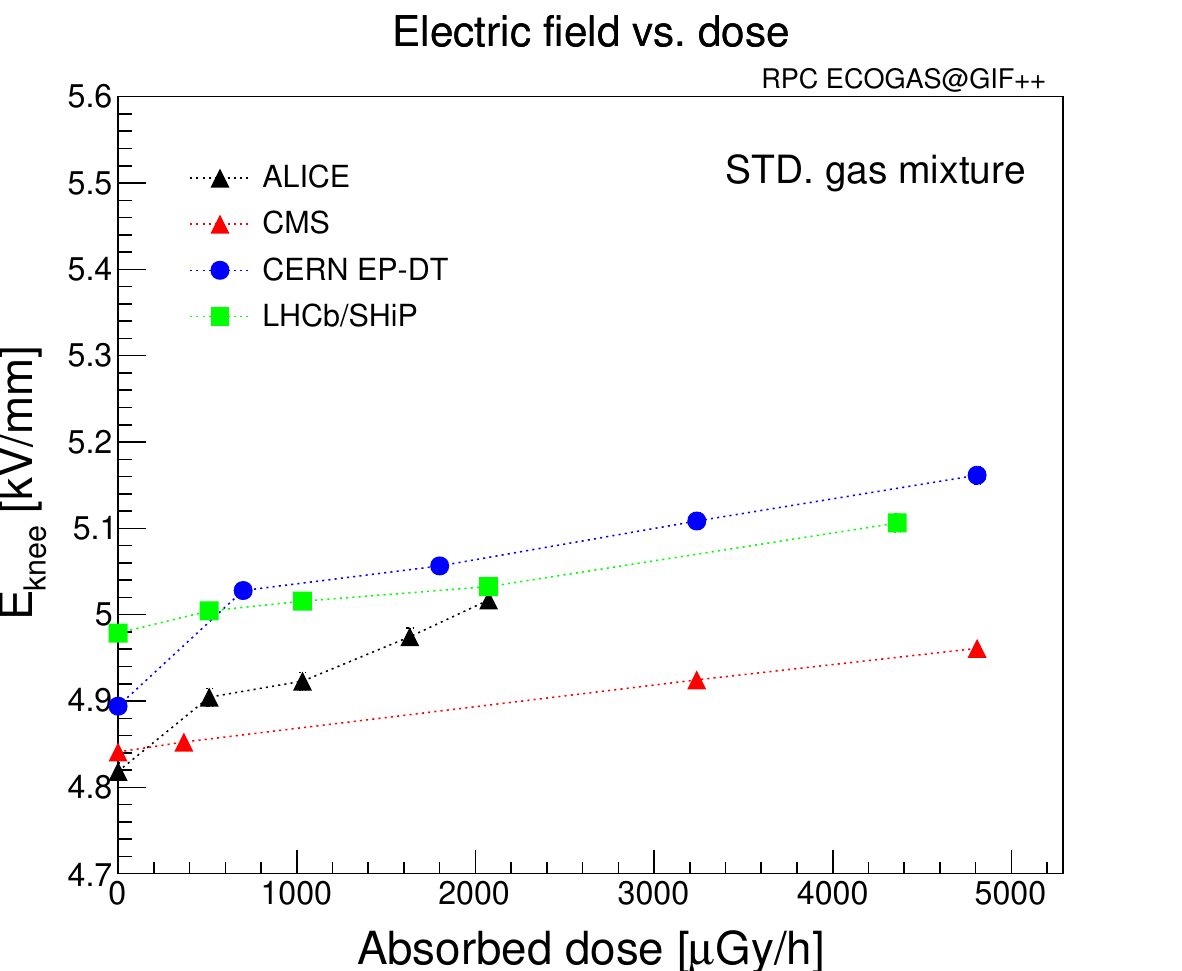}
\includegraphics[width=0.32\textwidth]{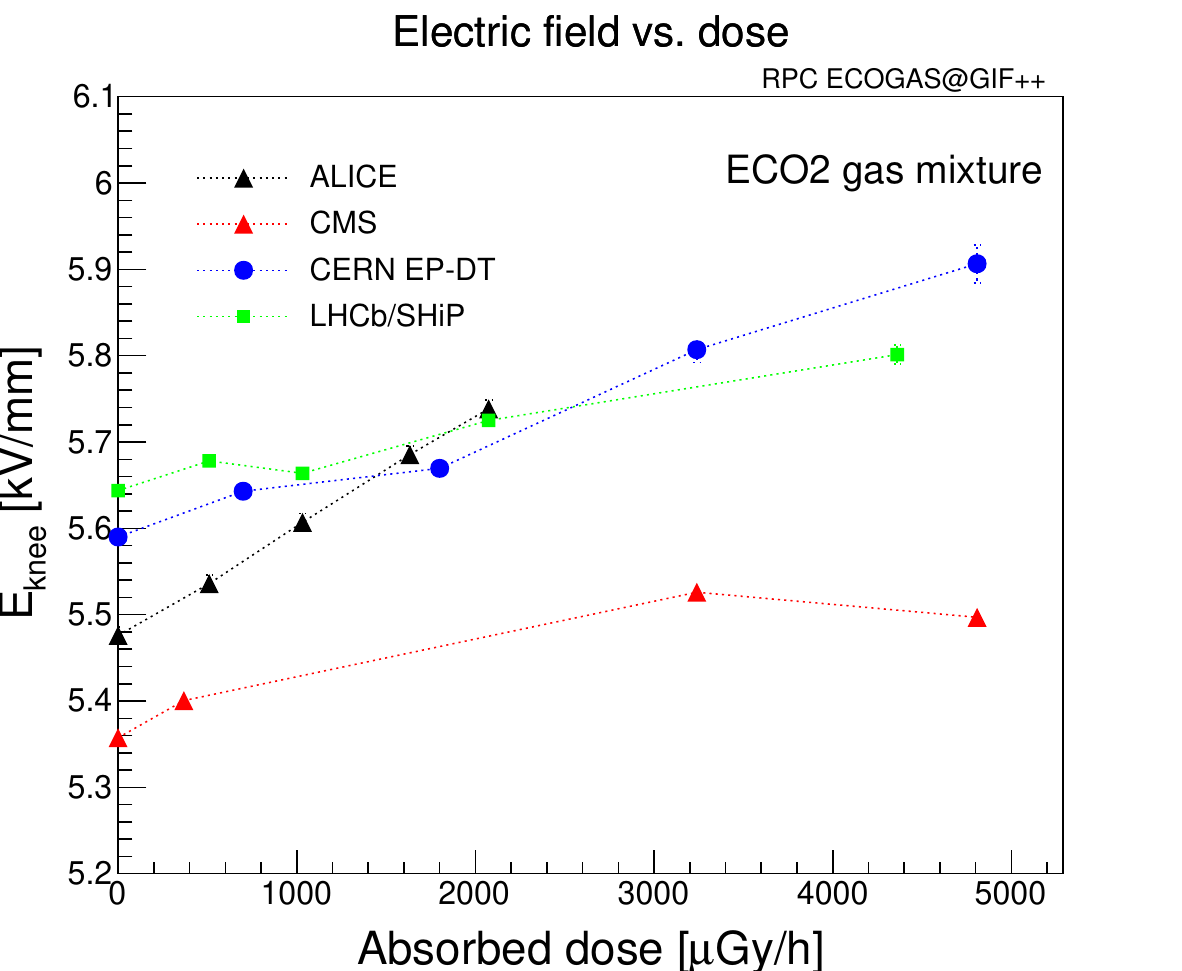}
\includegraphics[width=0.32\textwidth]{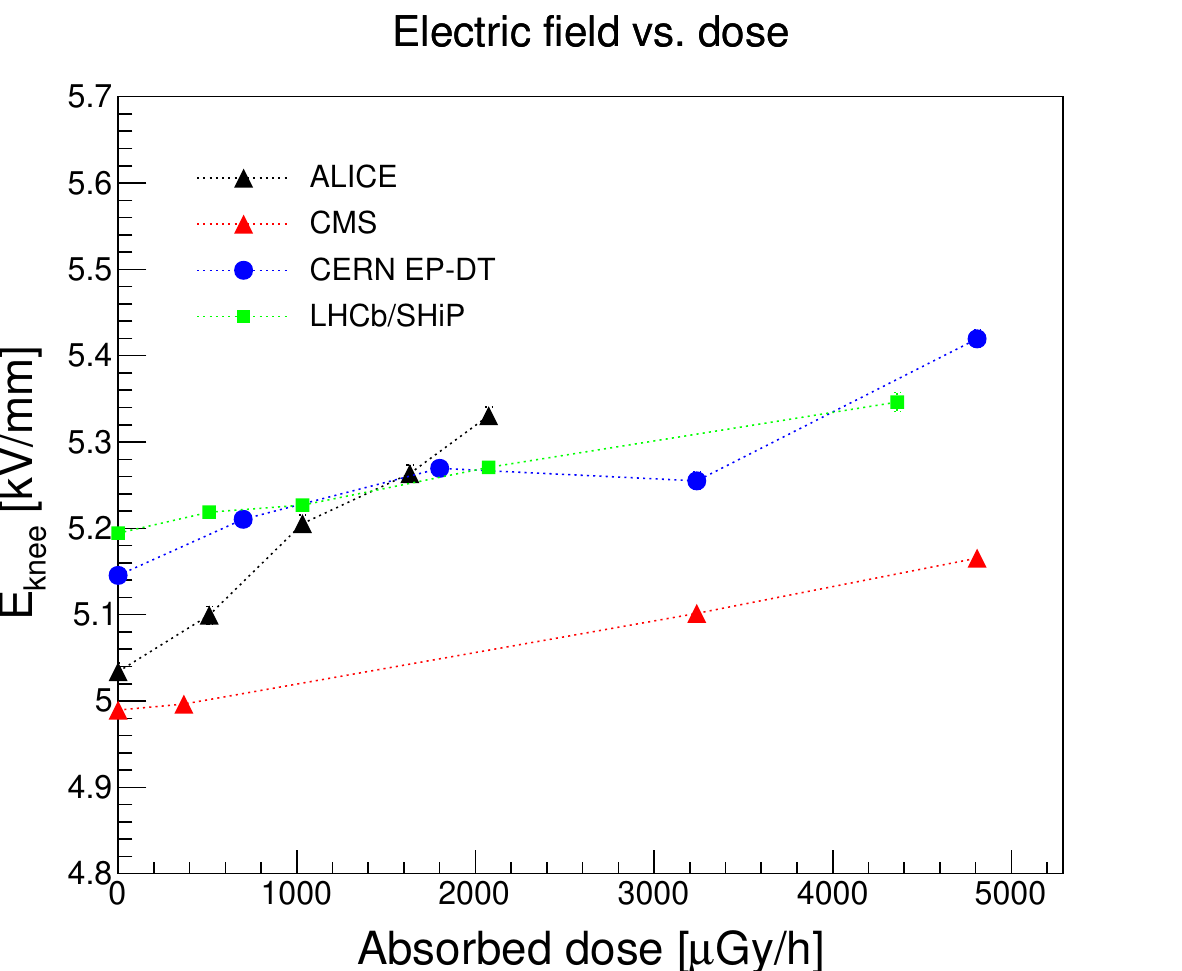}
\caption{Left: Electric field $E_\text{knee}$ at the knee, for the ALICE, CERN EP-DT, CMS and LHCb/SHiP chambers, as a function of the absorbed dose, when filled with STD mixture; Center: the same, but with the chambers filled with the ECO2 gas mixture; Right: the same, but with the chambers filled with the ECO3 gas mixture.}
\label{fig:EkneeDose}
\end{figure*}

The electric field $E_\text{knee}$ in the gas gap at the knee, for the four chambers under test, is shown in Figure~\ref{fig:EkneeDose} as a function of the absorbed dose. The left panel shows data obtained with chambers filled with the STD mixture, central and right panels with ECO2 and ECO3, respectively.

As expected, the shift of the efficiency curves toward higher operating voltages, observed in Figures ~\ref{fig:AliceHighRate} and ~\ref{fig:LHCbHighRate} at increasing irradiation, reflects itself in increased values of $E_\text{knee}$ at larger absorbed dose. This is a well known-effect, due to the intrinsic resistive nature of the electrodes in RPCs and the consequent voltage drop across them necessary to keep the current flowing from their outer to their inner surfaces. Basically, the voltage across the gas gap, $HV_\text{gas}$, is lower than the applied voltage, the difference generally computed, in first approximation, applying the Ohm's law. In the case of ECO2 and ECO3 mixtures, this drop is larger with respect to STD, due to the larger current densities involved. This might limit the chambers rate capability and, in fact, lower plateau efficiencies are observed with ECO2 and ECO3, as already pointed out and shown in Figure~\ref{fig:MaxEffDose}.

\begin{figure*}
\centering 
\includegraphics[width=0.32\textwidth]{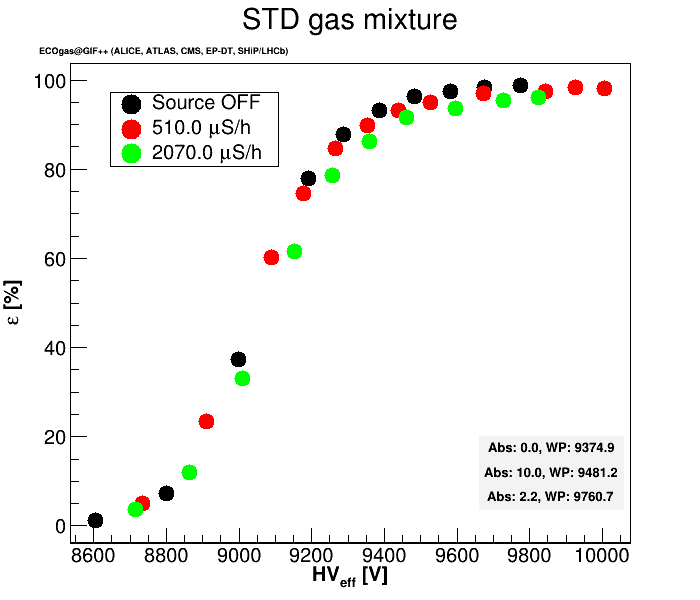}
\includegraphics[width=0.32\textwidth]{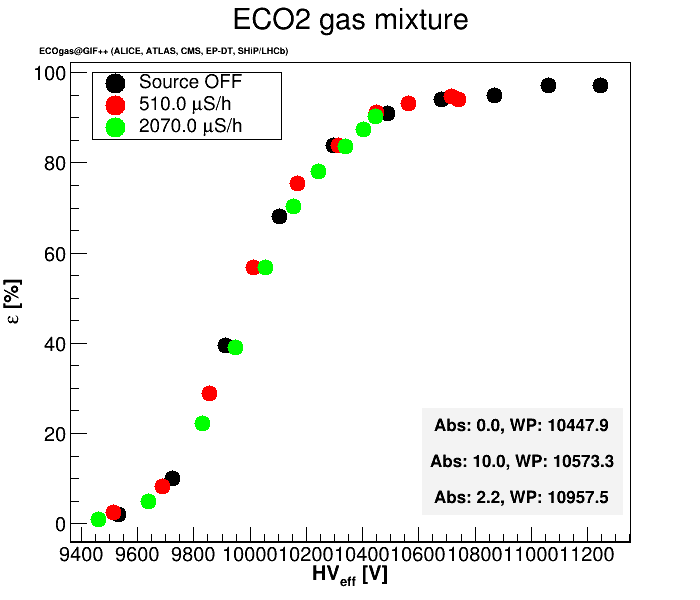}
\includegraphics[width=0.32\textwidth]{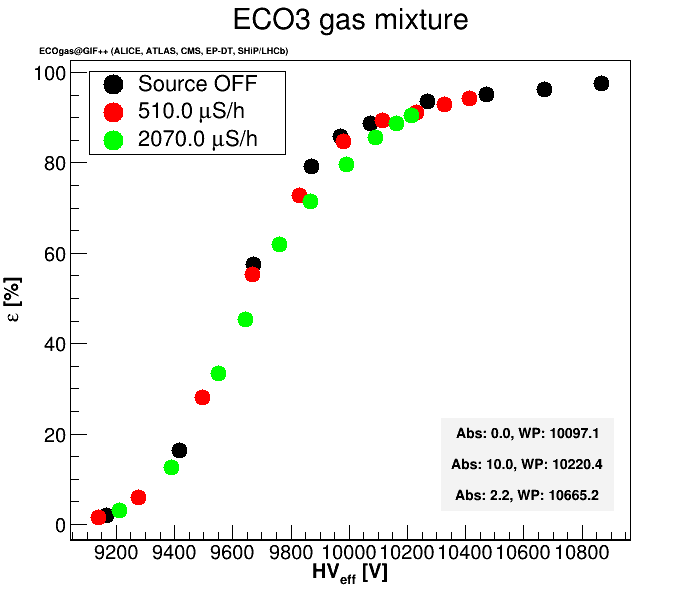}
\caption{Left: Efficiency and current density as a function of $HV_\text{gas}$, as defined in the text, for the ALICE chamber filled with the STD mixture, obtained with no irradiation, and with ABS= 10 and 22; Center: Same, but with the same chamber filled with the ECO2 mixture; Right: Same, but with the same chamber filled with the ECO3 mixture.}
\label{fig:AliceHVGas}
\end{figure*}

To check if the drop across the resistive electrode plates accounts for the observed features, the same efficiency curves of the ALICE chambers reported in Figure~\ref{fig:AliceHighRate} are plotted in Figure~\ref{fig:AliceHVGas}  as a function of the voltage drop across the gas gap, namely $HV_\text{gas} = HV_\text{eff} - RI = \rho J$, where $R$ is the total resistance across the HV electrodes, $I$ the current measured, and $\rho$ and $J$ are the bakelite resistivity and current density already introduced. Here a value for R $\approx 1 \times 10^7 \Omega$ was used, quite close to the resistance actually measured for the same chamber, in Autumn 2021, by filling it with Argon and measuring the slope of the current vs. applied voltage characteristic curve in its linear part. In fact, at high voltage, Argon behaves very much like a short circuit between the RPC electrodes. Indeed, as it can be seen from the figure, the curves roughly superimpose, supporting the hypothesis.

\begin{figure*}
\centering 
\includegraphics[width=0.32\textwidth]{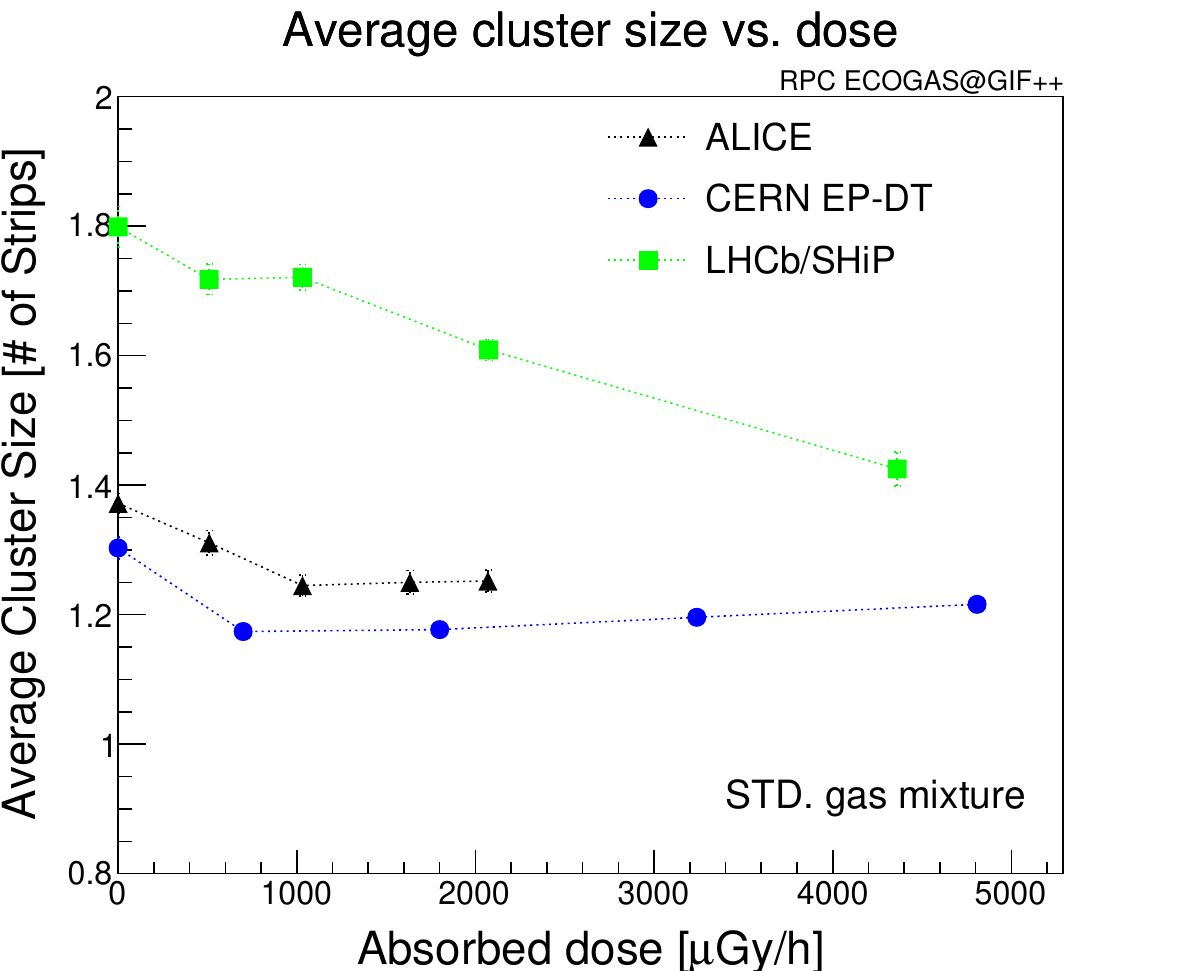}
\includegraphics[width=0.32\textwidth]{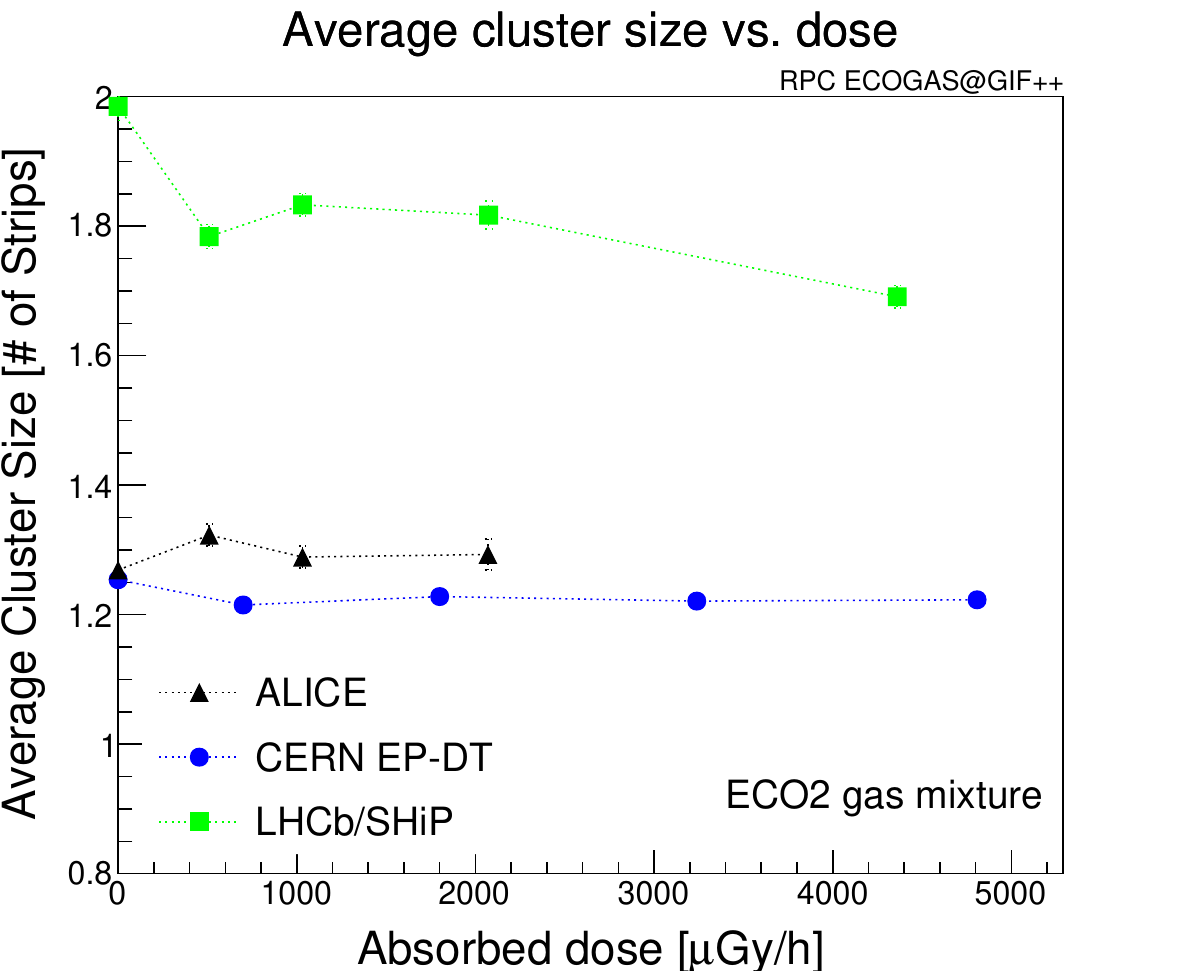}
\includegraphics[width=0.32\textwidth]{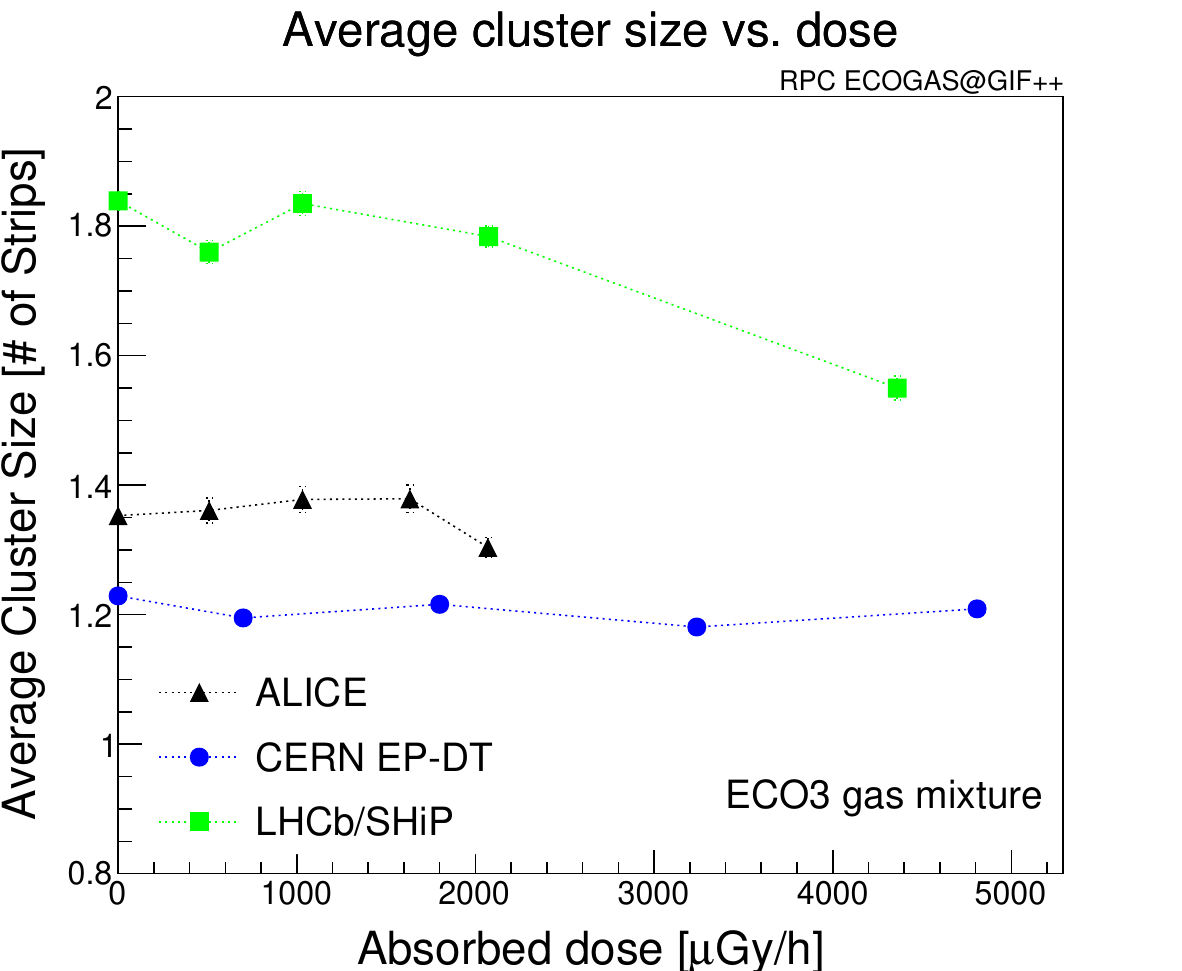}
\caption{Left: Average cluster size measured at $HV_\text{knee}$ and expressed in number of strips, vs. the absorbed dose, for the ALICE, CERN EP-DT, CMS and LHCb-SHiP chambers when filled with the STD gas mixture; Center: Same, but with the same chambers filled with the ECO2 mixture; Right: Same, but with the same chamber filled with the ECO3 mixture.}
\label{fig:ClusterVsDose}
\end{figure*}

The muon average cluster size, as already defined, was measured under irradiation from the $^{137}$Cs $\gamma$ source, and is shown in Figure~\ref{fig:ClusterVsDose}
for the detectors from ALICE, CERN EP-DT and LHCb/SHiP, and the three gas mixtures used as a function of the absorbed dose. Again, values reported in the plot were measured at $HV_\text{knee}$ and are expressed in number of strips. ALICE and CERN EP-DT chambers feature a cluster size between 1.2 and 1.4 strips, while the cluster size measured with the LHCb/SHiP reaches around 2, due the reduced strip pitch in this case. 

As a general trend, the average cluster size slightly decreases with increasing absorbed dose, following a behaviour similar to $\mathcal{E}_\text{max}$ and, due, essentially, to the slightly lower electric field at $HV_\text{knee}$ across the gas gap. However, even at high irradiation, muon average cluster size remains essentially comparable across STD, ECO2 and ECO3 mixtures, demonstrating that this is not a critical issue for what concerns the new eco-friendly gas mixtures, and spatial resolution stays approximately constant when passing from STD to ECO2 and ECO3 gas mixtures.

Similar considerations can be done about time resolutions, which, as already been pointed out, was not possible to precisely measure during this beam test, and will be the subject of a thorough study in the near future.

\begin{figure*}
\centering 
\includegraphics[width=0.32\textwidth]{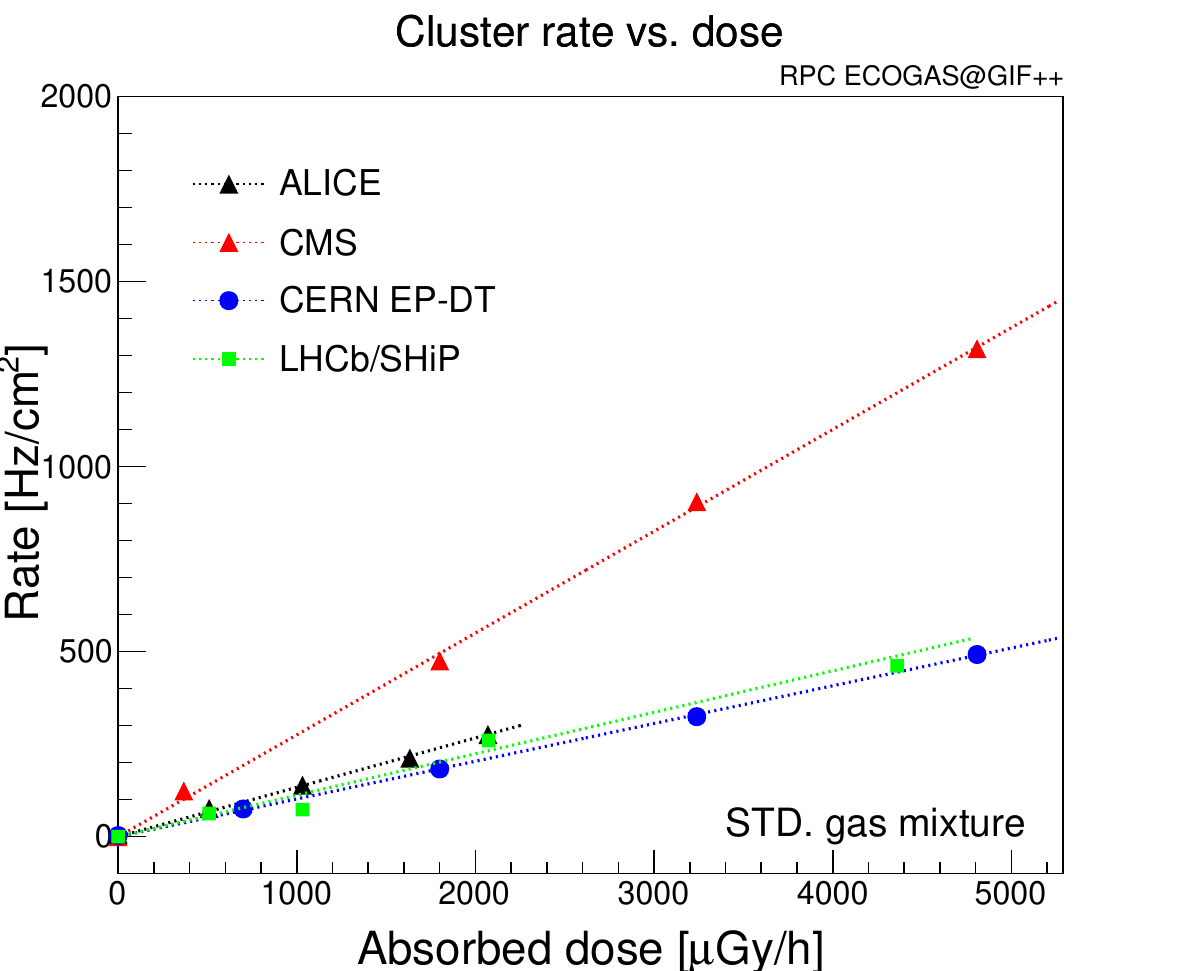}
\includegraphics[width=0.32\textwidth]{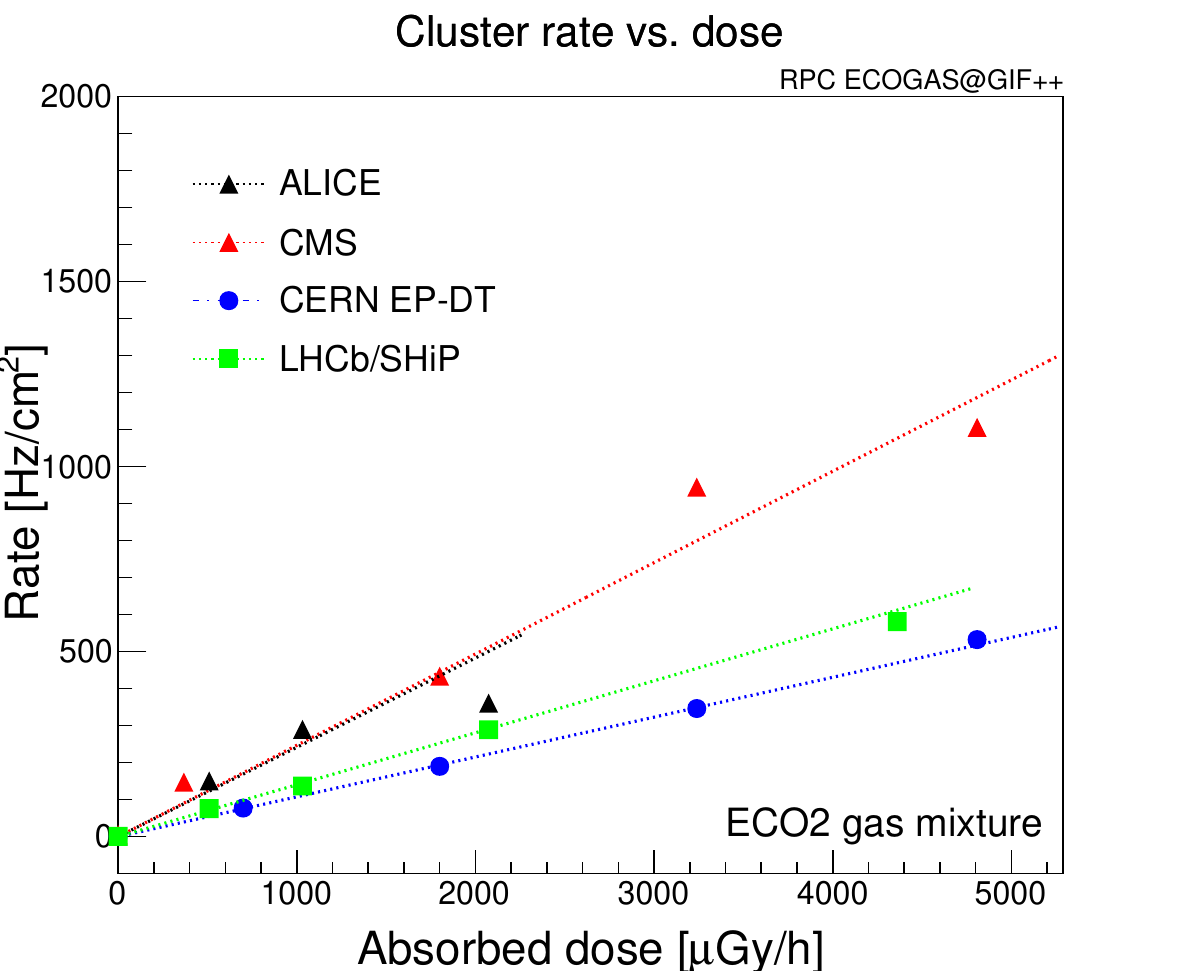}
\includegraphics[width=0.32\textwidth]{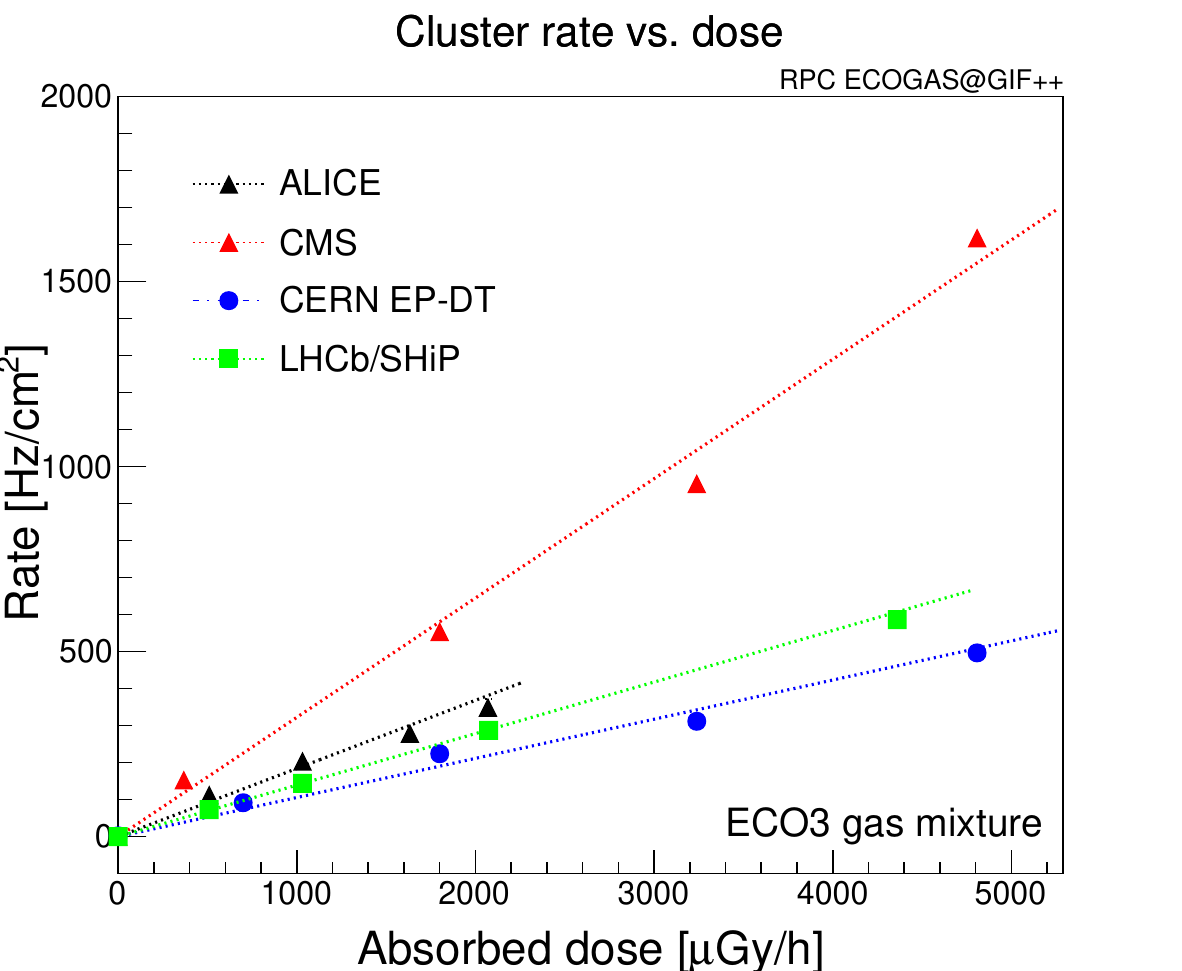}
\caption{Left: Cluster counting rate measured at $HV_\text{knee}$ vs. the absorbed dose, for the ALICE, CERN EP-DT, CMS and LHCb-SHiP chambers when filled with the STD gas mixture; Center: Same, but with the same chambers filled with the ECO2 mixture; Right: Same, but with the same chambers filled with the ECO3 mixture.}
\label{fig:RateVsDose}
\end{figure*}

The cluster $\gamma$ counting rate, namely the number of clusters related to $\gamma$ photons from the $^{137}$Cs per unit of area and unit of time, measured with the ALICE, CERN EP-DT, CMS and LHCb/SHiP chambers when operated at $HV_\text{knee}$, is shown in Figure~\ref{fig:RateVsDose} as a function of the absorbed dose, for the same values used for the plots shown in Figures ~\ref{fig:MaxEffDose}, ~\ref{fig:CurrDensDose}, ~\ref{fig:EkneeDose} and ~\ref{fig:ClusterVsDose}. Left panel refers to the chambers filled with the STD gas mixture, while central and right panels refer to ECO2 and ECO3. To be sure that these counts refer indeed to $\gamma$ photons, they were counted out of the muon spill time.

To guide the eye, a fit with a straight line constrained to pass through the origin is superimposed to the data points. Indeed, the expected direct proportionality of the cluster counting rate vs. absorbed dose is observed, with the largest fluctuations around this behaviour for the data taken with the chambers filled with the ECO2 mixture. Different slopes, namely different proportionality coefficients between cluster counting rate and dose are visible, and this implies that 
the measured counting rate, at the same ABS, changes depending on the chamber considered and the gas mixtures used to operate it. Actually, when inferring the counting rate from the absorbed dose, one has to consider the conversion probability of the impinging $\gamma$ in the detector,
whose exact value depends on the specific detector layout and materials, and the threshold of the electronics used. In addition, the doses measured at 3 and 6 m from the $^{137}$Cs source were considered and used in the graph, while the various chambers were not exactly at those distances from the source, as reported in Table \ref{tab:chamberdata}. Finally, note that the efficiency at $HV_\text{knee}$ of the various chambers, at different doses and different gas mixtures, is not the same, and all these factors should be taken into account when comparing the data shown in Figure~\ref{fig:RateVsDose}. 

The measured rates range from few hundreds Hz/cm$^2$ up to around 1.2 kHz/cm$^2$, featuring in the range of the expected backgrounds in the RPC systems of the ALICE, ATLAS and CMS experiments during the LHC-HL phase. For LHCb, where the use of RPCs in the future is still under discussion, rates might be much higher, and this demands for additional tests.

\section{Conclusions and Outlook}
\label{sec:conclusions}

The adoption of eco-friendly gas mixtures in RPCs presents a promising avenue for reducing the environmental impact and carbon footprint of high energy physics experiments. 

In this paper we have presented results obtained with eco-friendly gas mixtures, which could be used in the RPCs of the experiments at the LHC, replacing the one presently used, thereby significantly reducing the equivalent emission of CO$_2$ in the atmosphere. Focus was put on two mixtures, conventionally labelled as ECO2 and ECO3, whose main components are HFO-1234ze and CO$_2$ in various percentages (reported in Table \ref{tab:gasmixutre}), which substitutes TFE, being phased out because of its too large GWP. 

Experimental studies have shown that RPCs operated with these eco-friendly gas mixtures exhibit comparable detection efficiency, spatial and timing resolution to those using conventional gas mixtures. In particular we have proved that the ECO2 and ECO3 can be used for multiple RPC detector configurations and front-end electronics, operating with an efficiency larger than 90\% at a rate up to several hundreds Hz/cm$^2$.

However, the performance obtained with STD, ECO2 and ECO3 are not exactly superimposable. At similar irradiation and efficiency conditions, ECO2 and ECO3 exhibit larger current densities with respect to the STD mixture. This finds its counterpart in the measured charge distributions, characterized, for ECO2 and ECO3, by longer tails of large charge events. This leads to a decrease in efficiency with ECO2 and ECO3, up to around 10\%, at the highest rates tested here. For the rates expected during the HL-LHC phase in the RPC systems of the ALICE, ATLAS and CMS experiments, the efficiency drop ranges between 2 and 4\%.

This is not an issue for what concerns detection and trigger capabilities for the existing RPC systems of the LHC experiments, even during the High Luminosity phase, but it could be important for what concerns detector aging. As a matter of fact, aging processes are driven by the amount of current flowing through the electrodes and, more importantly, the gas. Gas pollutants produced in the avalanche and streamer processes can sometimes damage the inner electrode surfaces, producing local alterations in the electric field and/or points where pollutants accumulate more rapidly than elsewhere.

If the increase in the current density observed with ECO2 and ECO3 is really an issue in this respect is an open question. That is the reason why the test described in this paper must be complemented by long duration aging tests, which are the main task that the RPC ECOGas@GIF++ collaboration is presently taking care of. We plan to keep the chambers under test at the GIF++ for several years, collecting data all the way along, and continuously monitoring the performance of the detectors, to check if any deterioration is observed or not. The high irradiation conditions at GIF++ present the advantage to accelerate the aging processes with respect to what will happen during the HL-LHC phase, allowing to draw significant conclusions even after a limited (one-two years) period of time. We plan to continue these test even after that period, in order to prove, hopefully, that aging with these eco-friendly gas mixtures is at negligible, predictable, levels.  Results obtained on this important topic will be the subject of future publications.

Note that also other technical challenges need to be addressed, including the optimization of gas mixture composition, compatibility with RPC materials, and integration with existing systems. Therefore, continued research and development efforts are necessary to overcome these challenges, in order to ensure the successful implementation of eco-friendly gas mixtures in the RPC systems at the LHC experiments. 

Also, in principle, other gases may exist, with better properties than HFO-1234ze when used in gaseous detectors for particle physics, but whose usage is limited in industry and therefore whose procurement is difficult, if not virtually impossible. Nevertheless, the situation could change in the future, due to different requirements to the market. 

At another level, the search for eco-friendly gas mixtures in RPCs had also broader implications. It served as a catalyst for establishing a collaboration transversal to the LHC experiments, putting in common resources, fostering the exchange of ideas and sharing of best practises, experiences, and lessons learned. Through collaborative efforts and continuous innovation, the transition to eco-friendly gas mixtures in RPCs can serve as an exemplary model for sustainable scientific practices.

\section*{Data Availability Statement}
Data are available under reasonable request to the authors of this paper.

\end{document}